\documentclass[preprint2,twocolumn,times,tighten]{aastex631}
\usepackage{graphicx}	
\usepackage{amsmath}	
\usepackage{amssymb}	

\let\tablenum\relax
\usepackage{siunitx}

\begin{document}

\title{Reconnection-Driven Turbulent Fluctuations in the Magnetically Dominated Collisionless Regime}

\email{yuehu@ias.edu; *NASA Hubble Fellow}

\author[0000-0002-8455-0805]{Yue Hu*}
\affiliation{Institute for Advanced Study, 1 Einstein Drive, Princeton, NJ 08540, USA }

\author[0000-0001-8822-8031]{Luca Comisso}
\affiliation{Department of Astronomy and Columbia Astrophysics Laboratory, Columbia University, New York, NY, 10027, USA}

\author[0000-0002-1227-2754]{Lorenzo Sironi}
\affiliation{Department of Astronomy and Columbia Astrophysics Laboratory, Columbia University, New York, NY, 10027, USA}
\affiliation{Center for Computational Astrophysics, Flatiron Institute, 162 5th Avenue, New York, NY 10010, USA}

\author[0000-0002-0458-7828]{Siyao Xu}
\affiliation{Department of Physics, University of Florida, 2001 Museum Rd., Gainesville, FL 32611, USA}



\begin{abstract}
Magnetic reconnection is a fundamental plasma process that converts magnetic energy into bulk flow energy, thermal energy, and nonthermal particle acceleration. Despite its importance, the statistical properties of the turbulent fluctuations generated by collisionless reconnection, which are essential for understanding how this energy conversion proceeds, remain poorly understood. Here, we employ large-scale 3D particle-in-cell simulations to investigate the turbulence characteristics of velocity and magnetic field fluctuations generated by collisionless reconnection in a magnetically dominated pair plasma. We characterize their statistical properties by computing structure functions along different directions within the reconnection layer. We find that the square root of the second-order velocity structure function follows a power-law scaling with a slope $\sim1/3$ at intermediate to large scales. The square root of the second-order magnetic structure function consistently exhibits a steeper slope, in the range $\sim 0.6 - 0.8$. The presence of a finite guide field does not systematically modify the slope of the velocity fluctuations, while it progressively steepens the scaling of the magnetic fluctuations in the guide-field and inflow directions. We measure higher-order structure functions, which reveal strong magnetic intermittency along the outflow direction and weaker intermittency in the inflow and guide-field directions. Additionally, the local anisotropies of both velocity and magnetic field fluctuations are greater for stronger guide fields. These results provide a systematic characterization of the multiscale nature of turbulence in collisionless and magnetically dominated reconnection layers, with important implications for plasma heating and particle acceleration.
\end{abstract}




\section{Introduction} \label{sec:intro}

Magnetic reconnection reconfigures the magnetic‑flux topology, releasing magnetic energy into bulk plasma flows, thermal heating, and non-thermal particles \citep{1957JGR....62..509P,1958Obs....78...30S,1964NASSP..50..425P,LV99,2020PhPl...27a2305L,2022NatRP...4..263J,2025ARA&A..63..127S}. Wherever reconnection occurs—whether in solar flares \citep{1976SoPh...50...85K,1994Natur.371..495M}, magnetospheric substorms \citep{1998JGR...103.4419N,2008Sci...321..931A}, pulsar‑wind nebulae \citep{2001ApJ...547..437L,Comisso2020}, or accretion‑disk coronae \citep{1990A&A...227..473A,Beloborodov17}—it is invariably accompanied by a cascade of fluctuations that originate inside the reconnecting layer itself \citep{2011NatPh...7..539D,2015ApJ...806L..12O,2016ApJ...818...20H,2017ApJ...838...91K,2019PhPl...26g2121S,2021ApJ...919..111G,2024PhPl...31h2119H}. 
Quantifying the statistical properties of reconnection‑driven fluctuations is therefore essential for understanding how magnetic energy is released and partitioned within the plasma and for building predictive models of particle acceleration and transport.

\begin{figure*}[htbp!]
  \centering
  \gridline{
    \fig{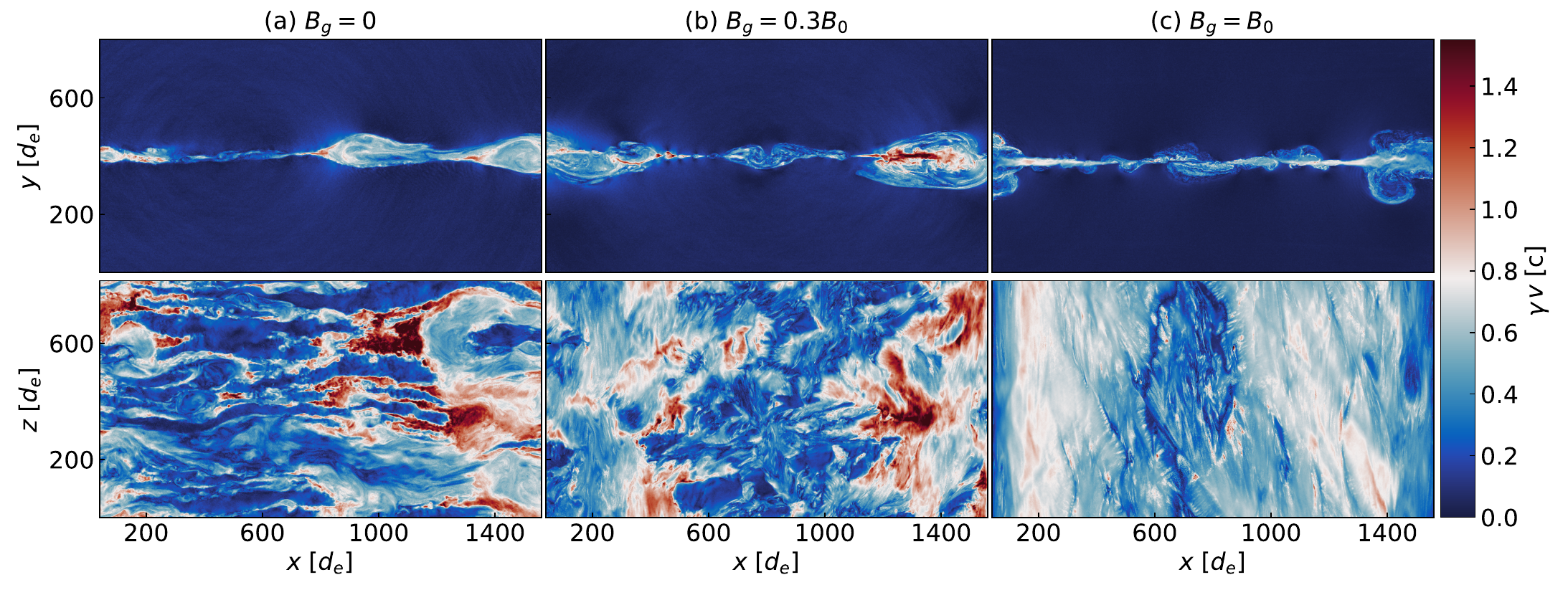}{1\textwidth}{}
  }
  \vspace{-4em}
  \gridline{
    \fig{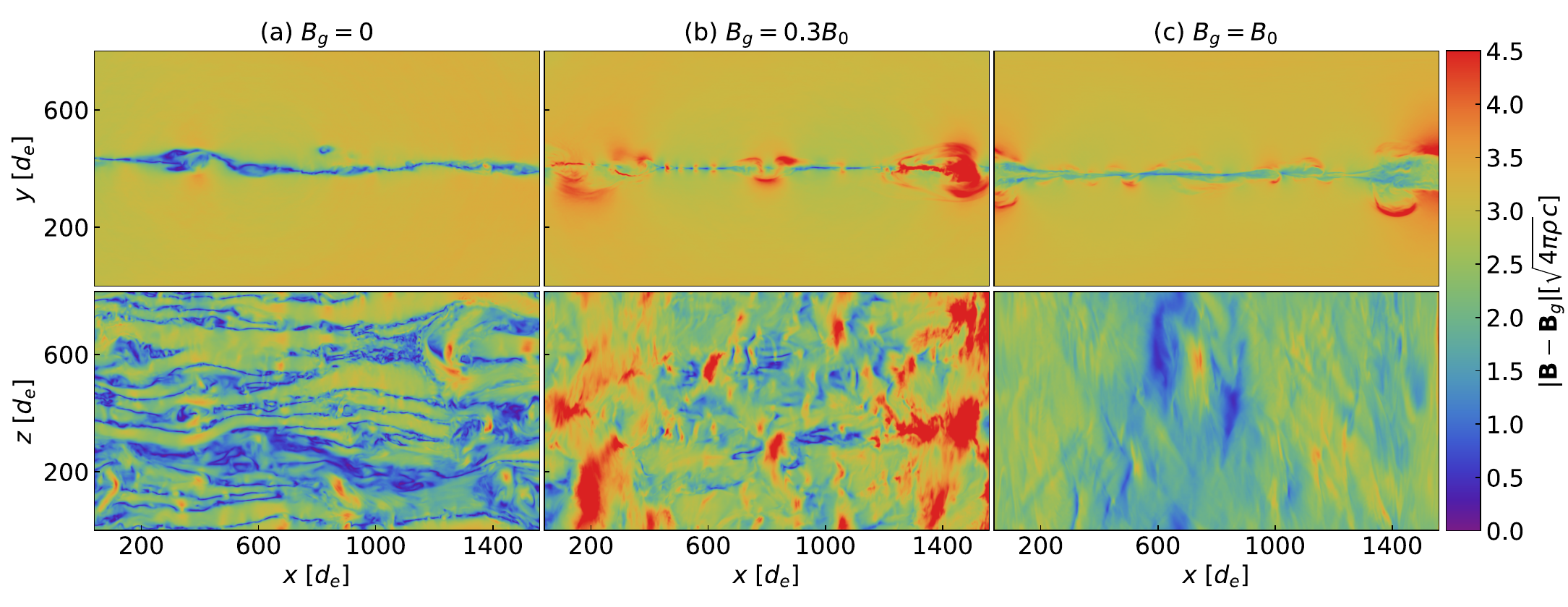}{1\textwidth}{}
  }
\vspace{-2.5em}
  \caption{Velocity slices (top two rows) and magnetic field slices (bottom two rows), respectively, taken at the $z$ location where the flux ropes are visually most prominent, and taken at the center of the $y$ axis. The left panels do not include a guide field, whereas the middle and right panels include a guide field of strength $B_g=0.3B_0$ or $B_g=B_0$ (oriented along the $z$-axis), respectively. $\rho=n_0m$ is the mass density and $\pmb{B}_g=B_g\hat{z}$.}
  \label{fig:maps}
\end{figure*}

Within the magnetohydrodynamic (MHD) approximation, numerical studies have started to probe these statistical properties. An initially laminar current sheet can self-generate turbulent fluctuations with a Kolmogorov-like inertial range and scale-dependent anisotropy once reconnection enters a quasi-steady state, yielding reconnection rates that are largely insensitive to explicit resistivity \citep{2017ApJ...838...91K,2017ApJ...834...47B,2025arXiv251001060V,2025arXiv251009978L}.

Nevertheless, many astrophysical plasmas are collisionless or weakly collisional, such that kinetic effects—e.g., finite-Larmor-radius physics and collisionless dissipation—become essential and cannot be captured within the MHD framework \citep{2001ApJ...562L.129N,2009ApJS..182..310S,2019LRSP...16....5V}. Fully kinetic particle-in-cell (PIC; \citealt{BL85}) simulations therefore provide a necessary tool to access these regimes and have revealed a rich, multiscale reconnection process with efficient non-thermal particle acceleration \citep{2014ApJ...783L..21S,2017PhPl...24i2110D,LiX2019,Comisso2018,Comisso2019,2021ApJ...919..111G,2021JPlPh..87c9028A,zhang_sironi_21,Comisso2022,zhang_23}. Within this broad class of collisionless plasmas, an important subset corresponds to relativistic, magnetically dominated systems, in which the magnetization parameter $\sigma \gtrsim 1$. Such conditions are expected in environments including pulsar magnetospheres and winds, magnetar magnetospheres, and black-hole magnetospheres, as well as in magnetically dominated relativistic jets \citep{1969ApJ...157..869G,1977MNRAS.179..433B,1996AIPC..366..111D,2009MNRAS.394.1182K,2012SSRv..173..341A,2014ApJ...783L..21S,2017ARA&A..55..261K,2025ARA&A..63..127S}. These systems provide a particularly clean setting to study reconnection-driven turbulence in the high-$\sigma$ regime. Nevertheless, a comprehensive statistical characterization of reconnection-driven turbulent fluctuations in the magnetically dominated collisionless regime is still lacking.


Here we present the results of large-scale three-dimensional (3D) PIC simulations of magnetic reconnection in magnetically-dominated plasmas. We follow the evolution of an initially laminar current sheet through the onset of collisionless reconnection into a state of reconnection-driven turbulence. By resolving scales from the system size down to the electron inertial length, we characterize the statistical properties of velocity and magnetic field fluctuations, compute their intermittency and anisotropy, and assess the influence of the guide field on the cascade. The resulting picture links the reconnection dynamics to the associated energy transfer across multiple scales, providing a basis for understanding reconnection-driven turbulent fluctuations in the regime of magnetically dominated collisionless plasma.


\section{Numerical simulations} 
\label{sec:data}
We analyze large-scale 3D PIC simulations performed with TRISTAN-MP \citep{buneman_93, spitkovsky_05}. 
We consider a cold electron-positron plasma upstream of the current sheet, with rest-frame density $n_0$ represented by two computational particles per cell. The cold upstream pair plasma has $k_BT/mc^2=10^{-4}$. The hot current sheet is initialized with an overdensity of $3$ as compared to the upstream, and a temperature of $k_BT_h/mc^2=\sigma/6$ to maintain pressure balance. The hot plasma carries the electric current via a small $\pm \hat{z}$ drift of electrons/positrons. The reconnecting field is initialized in a Harris equilibrium with
$\mathbf{B}=-B_{0}\hat{x}\tanh(2\pi y/\Delta)$, with sheet thickness $\Delta\simeq 50\, d_e$, where $d_e=c/\omega_{\rm p}$ is the plasma skin depth. The strength $B_0$ of the reversing field is parameterized by the magnetization $\sigma = B_0^2 / 4\pi  n_0 m c^2 = \left(\omega_{\rm c} / \omega_{\rm p}\right)^2$, where $\omega_{\rm c} = e B_0 / m c$ is the gyrofrequency and $\omega_{\rm p} = \sqrt{4\pi n_0 e^2 / m}$ is the plasma frequency. We consider $\sigma=10$ as a representative case of the  $ \sigma\gg1$ regime of ``relativistic reconnection'' \citep{sironi_review}. We compare simulations with different values of a uniform guide field initialized along $z$, with strengths $B_g = (0, 0.1, 0.3, 0.5, 1) B_0$.  The numerical speed of light is $c=0.45$ cells per timestep, which resolves the Larmor period of upstream particles with 11 timesteps. 
To reduce numerical noise, the electric current is filtered at each step with 16 passes of a 1-2-1 digital filter along each axis.

Along the inflow ($y$) direction, two injectors continuously introduce fresh plasma and magnetic flux into the domain, while receding away from the midplane \citep{sironi_16}. We employ periodic boundary conditions in $z$ and outflow boundaries in $x$. We resolve the plasma skin depth $d_e=c/\omega_{\rm p}$ with 2.5 cells. 
The grid-based variables saved at output times are downsampled by a factor of 5 in each direction, so the smallest length scale probed by our output variables is $2\,d_e$. We employ large domains, adopting $L=L_z=800 \, d_e$.  Here, $L$ denotes the half-length of the domain along the outflow direction $x$, while $L_z$ is the full length in the $z$ direction. These large domains are essential to capture distinctive 3D effects \citep[e.g.,][]{zhang_sironi_21,zhang_23}. Throughout the analysis, all lengths are expressed in units of $d_e$. Future work should assess the degree of self-similarity of these results, namely how the turbulent fluctuations in reconnection depend on the scale separation between $L$ and $d_e$.

We present results at $t=5.625L/c$ for all guide-field cases, when the reconnection dynamics is in a statistical steady state in which the reconnection rate is approximately time-independent (see Fig.~\ref{fig:beta_rec} in the Appendix), and the fluctuation statistics (as quantified by the fluctuation's energy) are stationary (see Fig.~\ref{fig:EkEb} in the Appendix). We verified that the main conclusions remain robust at other steady-state times (see Figs.~\ref{fig:maps_004}, \ref{fig:maps_006}, \ref{fig:sf_time_004}, \ref{fig:sf_time_006}, \ref{fig:sf_slope_004}, \ref{fig:sf_slope_006},  \ref{fig:sf_nth_004}, and \ref{fig:sf_nth_006} in Appendix for $t=4.5 L/c$ and $t=6.75L/c$). We adopt absorbing boundaries in $x$, as periodic $x$ boundaries lead to accumulation of particles and magnetic flux that eventually suppresses reconnection globally \citep{sironi_16}. Since different boundary conditions can alter the reconnection dynamics throughout the entire domain, a direct statistical comparison of fluctuation properties between the two boundary treatments is warranted but is beyond the scope of this study.

\section{Results} 
\label{sec:results}
\subsection{Velocity and magnetic field fluctuations driven by reconnection}
Fig.~\ref{fig:maps} shows 2D slices of the velocity and magnetic fluctuations in simulations with guide-field strengths $B_g=0$, $0.3B_0$, and $B_0$. The $x-y$ slices are taken at $z$ location where the flux ropes are visually most prominent, while the $x-z$ slices are taken in the middle of the $y$ axis. In all cases, the velocity and magnetic fluctuations are generated self-consistently by magnetic reconnection in the collisionless, magnetically dominated regime, extending down to the skin-depth scale. These reconnection-driven fluctuations were also observed in earlier MHD work at larger scales \citep{2017ApJ...838...91K,2017ApJ...834...47B,2025arXiv251001060V,2025arXiv251009978L}. Without a guide field, large flux-rope structures form within the current sheet or subsequently merge or fragment, driving multiscale velocity fluctuations elongated in the outflow ($x$) direction. When a moderate guide field is present ($B_g=0.3B_0$), the velocity fluctuations exhibit elongation both along the outflow direction and along the guide-field direction. For a strong guide field ($B_g=B_0$), the velocity fluctuations become preferentially stretched along the $z$ direction, reflecting the dominant role of the guide field in establishing some degree of invariance along the $z$ direction.

A similar trend is seen in the magnetic-field fluctuations. In the $B_g=0$ case, the magnetic field forms large flux ropes within the current sheet that undergo bending and distortion. These kink-like deformations are most visible in the $x-z$ plane, where the absence of guide-field tension allows the flux ropes to bend in $z$. With a moderate guide field ($B_g=0.3B_0$), such deformations are partially suppressed, yielding structures elongated along $x$ and $z$. For a strong guide field ($B_g=B_0$), the magnetic morphology becomes significantly more coherent, with structures preferentially elongated along the guide-field direction and largely stabilized against kink-type distortions.

\begin{figure*}
\centering
\includegraphics[width=1.0\linewidth]{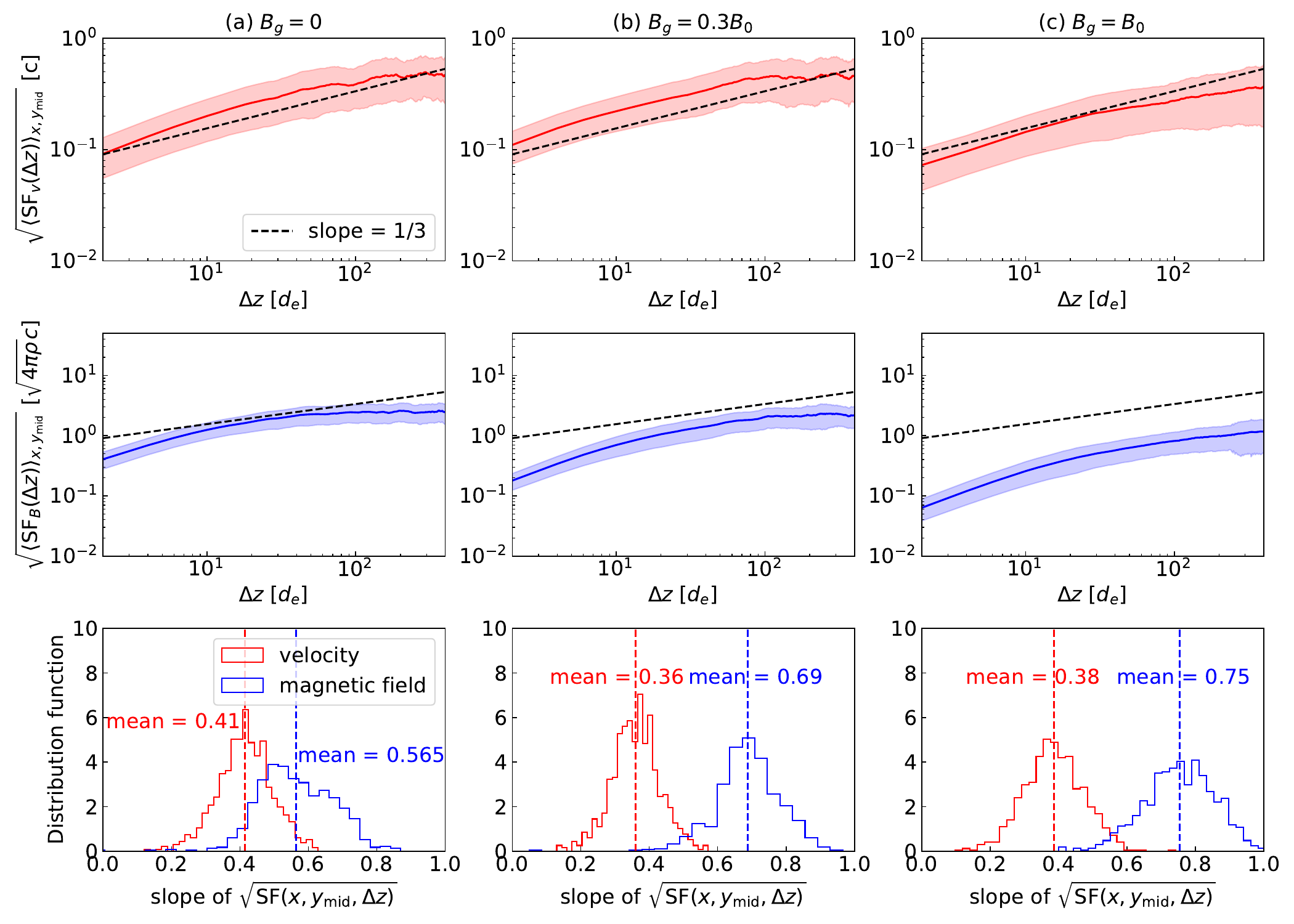}
    \caption{\textbf{Top and middle panels:} Square root of the second-order structure function (SF) for the velocity (red) and magnetic field (blue). The SF is computed for separations along the $z$-axis at each position $(x,y_{\rm mid})$, where $y_{\rm mid}$ denotes the midpoint in $y$ for each $x$ within the reconnection region. The resulting ${\rm SF}(x, y_{{\rm mid}}, \Delta z)$ are then averaged over all $x$ values, denoted as $\langle{\rm SF}(\Delta z)\rangle_{x,y_{\rm mid}}$. The shaded area represents the standard deviation calculated from the data points at a given spatial separation. The black dashed line indicates the slope of 1/3. Left panels show results without a guide magnetic field; the middle and right panels include guide fields $B_g=0.3B_0$ and $B_g=B_0$, respectively. \textbf{Bottom panel:} The distribution of the $\sqrt{{\rm SF}(x, y_{\rm mid},\Delta z)}$'s slope. The slope is fitted over separations in the range 2 – 30$d_e$. The red and blue dashed lines represent the mean slopes of the velocity and magnetic field structure functions, respectively.}
    \label{fig:sf_xmid}
\end{figure*}

\subsection{Statistical characterization via second order structure functions}
To quantitatively characterize the reconnection-driven fluctuations, we compute the second-order structure functions of velocity and magnetic field along the $x$, $y$, and $z$ directions, focusing on the reconnection region, Here, we define the “reconnection region” as the volume satisfying a criterion based on a mixing factor \citep{Daughton14,Rowan17,zhang_sironi_21} between the two oppositely-directed inflows toward the midplane. Particles originating from above and below the midplane are classified as two distinct populations, and the reconnection region is defined as the volume where both populations contribute at least 15\% to the local plasma density. Tests of using thresholds of 10\% and 20\% are given in the Appendix's Figs.~\ref{fig:sf_slope_005_th10} and \ref{fig:sf_slope_005_th20}. 

For a general vector field $\pmb{f}(x,y,z)$ (either the velocity $\pmb{v}$ or magnetic field $\pmb{B}$) within the reconnection region, we define:
\begin{equation}
\begin{aligned}
\label{eq.sf}
{\rm SF}(x,y,\Delta z)=\langle|\pmb{f}(x,y,z)-\pmb{f}(x,y,z+\Delta z)|^2\rangle,\\
{\rm SF}(x,\Delta y,z)=\langle|\pmb{f}(x,y,z)-\pmb{f}(x,y+\Delta y,z)|^2\rangle,\\
{\rm SF}(\Delta x,y,z)=\langle|\pmb{f}(x,y,z)-\pmb{f}(x+\Delta x,y,z)|^2\rangle,
\end{aligned}
\end{equation}
where $\langle\cdot\rangle$ denotes an average over all point‑pairs separated by the indicated separation within the reconnection region. We employ linear binning with a bin width of one output cell, rounding separation distances to the nearest integer. The analysis uses uniform weighting, in which the sum of squared increments in each bin is normalized to the total count of valid pairs found at that separation. The shaded regions represent the standard deviation of the structure function values calculated within each bin at the corresponding scale. Since the statistics of fluctuations may differ along each axis (plasma inflow along $y$, reconnection outflow along $x$, and electric current/guide field along $z$), these structure functions probe distinct contributions to the overall velocity or magnetic fluctuations. We use the subscripts ``$v$'' and ``$B$'' to distinguish the velocity structure function ${\rm SF}_v$ and magnetic field structure function ${\rm SF}_B$.

\subsubsection{Out of plane ($z$) direction}
Fig.~\ref{fig:sf_xmid} shows the square root of the structure function, $\sqrt{\langle{\rm SF}(\Delta z)\rangle_{x,y_{\rm mid}}}$, where $\langle{\rm SF}(\Delta z)\rangle_{x,y_{\rm mid}}$ is computed for separations along $z$ at each position $(x,y_{\rm mid})$ and subsequently averaged over all $x$. Here, $y_{\rm mid}$ denotes the midpoint in $y$ for each fixed $x$ within the reconnection region. The slope (of the square root of the structure function) is equivalent to the structure function's slope divided by two. The left panel of Fig.~\ref{fig:sf_xmid} corresponds to the case without a guide magnetic field, while the middle and right panels include a guide field, $B_g=0.3B_0$ and $B_g=B_0$, respectively.

In the absence of a guide field, $\sqrt{\langle{\rm SF}_v(\Delta z)\rangle_{x,y_{\rm mid}}}$ follows an approximate power law with slope $\sim 1/3$ at intermediate scales ($\sim10 –100d_e$), and transitions to a steeper scaling at smaller separations. Introducing a guide field ($B_g=0.3B_0$ or $B_g=B_0$) does not appreciably change the slope, aside from a marginal decrease in amplitude for $B_g=B_0$. In the strong guide-field case ($B_g=B_0$), the slope is slightly steeper than 1/3 at small scales ($\lesssim10d_e$), whereas at larger scales it is marginally shallower than 1/3. The flattening of $\sqrt{\langle{\rm SF}_v(\Delta z)\rangle_{x,y_{\rm mid}}}$ approximately starts from $100d_e$, corresponding to the largest injection scale and the characteristic thickness of flux ropes. The inferred injection scale is consistent at $t=4.5L/c$ and $t=6.75L/c$ (see Figs.~\ref{fig:sf_time_004} and \ref{fig:maps_006} in the Appendix), indicating that the effective layer thickness has stabilized over the interval used to compute the reported statistics.

For the magnetic field, $\sqrt{\langle{\rm SF}_B(\Delta z)\rangle_{x,y_{\rm mid}}}$ displays a different behavior. Without a guide field, the square root of the second-order magnetic structure function is steeper than its velocity counterpart at scales $\lesssim 30 d_e$, while at larger separations it gradually flattens to slopes shallower than 1/3. With a guide field ($B_g=0.3B_0$ or $B_g=B_0$), the magnetic structure functions steepen further below $\sim30d_e$, and their amplitude decreases systematically. As visible in Fig.~\ref{fig:maps}, the presence of a significant guide field modifies the morphology of flux ropes, and the shift toward larger-scale magnetic structures is consistent with the observed trend of the magnetic fluctuations in Fig.~\ref{fig:sf_xmid}. Overall, the amplitude of $\sqrt{\langle{\rm SF}_B(\Delta z)\rangle_{x,y_{\rm mid}}}$ decreases in the presence of a guide field, with the most pronounced reduction occurring for $B_g=B_0$. These statistics, for both velocity and magnetic field, are stable in the steady state, as shown in the Appendix's Figs.~\ref{fig:sf_time_004} and \ref{fig:sf_time_006}.

To examine how the structure-function slope varies at different positions in the layer, we fit $\sqrt{{\rm SF}(x,y_{\rm mid}, \Delta z)}$ in the range of separations between 2 and 30$d_e$. The slope distributions are shown in the bottom panels of Fig.~\ref{fig:sf_xmid}. The slopes of the square-root velocity structure functions vary widely, from nearly flat ($\sim0.1$) to relatively steep ($\sim0.7$), but the mean remains close to 1/3 over the available range of scales.
However, the slopes of the square-root magnetic structure functions are systematically steeper, reaching values up to $\sim$1.0. The mean slopes increase from 0.565 for $B_g=0$, 0.69 for $B_g=0.3B_0$, and further to 0.75 for $B_g=B_0$.

\begin{figure*}
\centering
\includegraphics[width=1.0\linewidth]{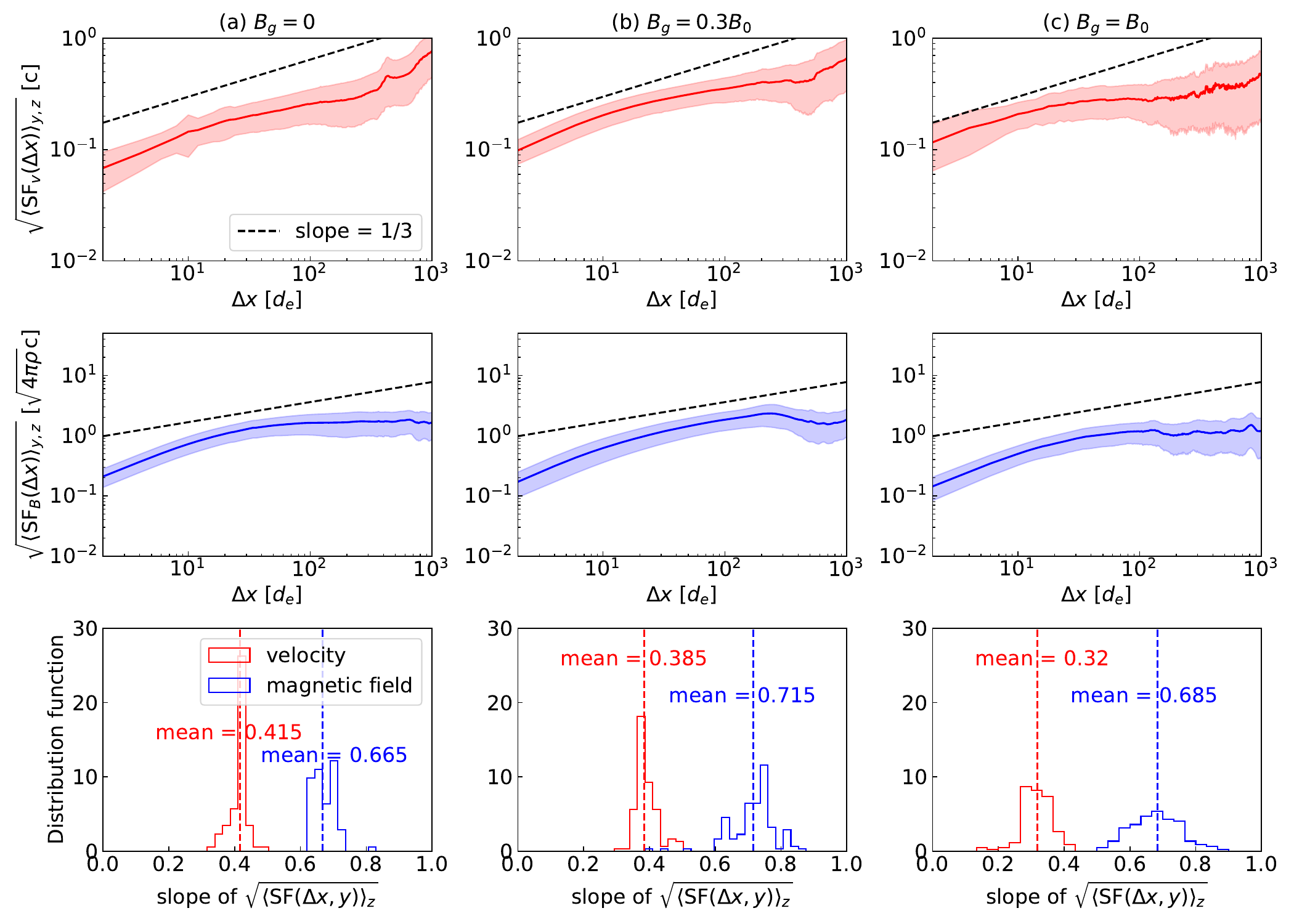}
        \caption{\textbf{Top and middle panels:} Square root of the second-order structure function (SF) for the velocity (red) and magnetic field (blue). The SF is computed for separations along the $x$-axis at each position $(y, z)$. The resulting ${\rm SF}(\Delta x,y,z)$ are then averaged over all $y$ and $z$ values, denoted as $\langle{\rm SF}(\Delta x)\rangle_{y,z}$. The shaded area represents the standard deviation calculated from the data points at a given spatial separation. The black dashed line indicates the slope of 1/3. Left panels show results without a guide magnetic field; the middle and right panels include guide fields $B_g=0.3B_0$ and $B_g=B_0$, respectively. \textbf{Bottom panel:} The distribution of the $\langle{\rm SF}(\Delta x,y)\rangle_z$'s slope. The slope is fitted over separations in the range 2 – 30$d_e$. The red and blue dashed lines represent the mean slopes of the velocity and magnetic field structure functions, respectively.}
    \label{fig:sf_dy}
\end{figure*}

\begin{figure*}
\centering
\includegraphics[width=1.0\linewidth]{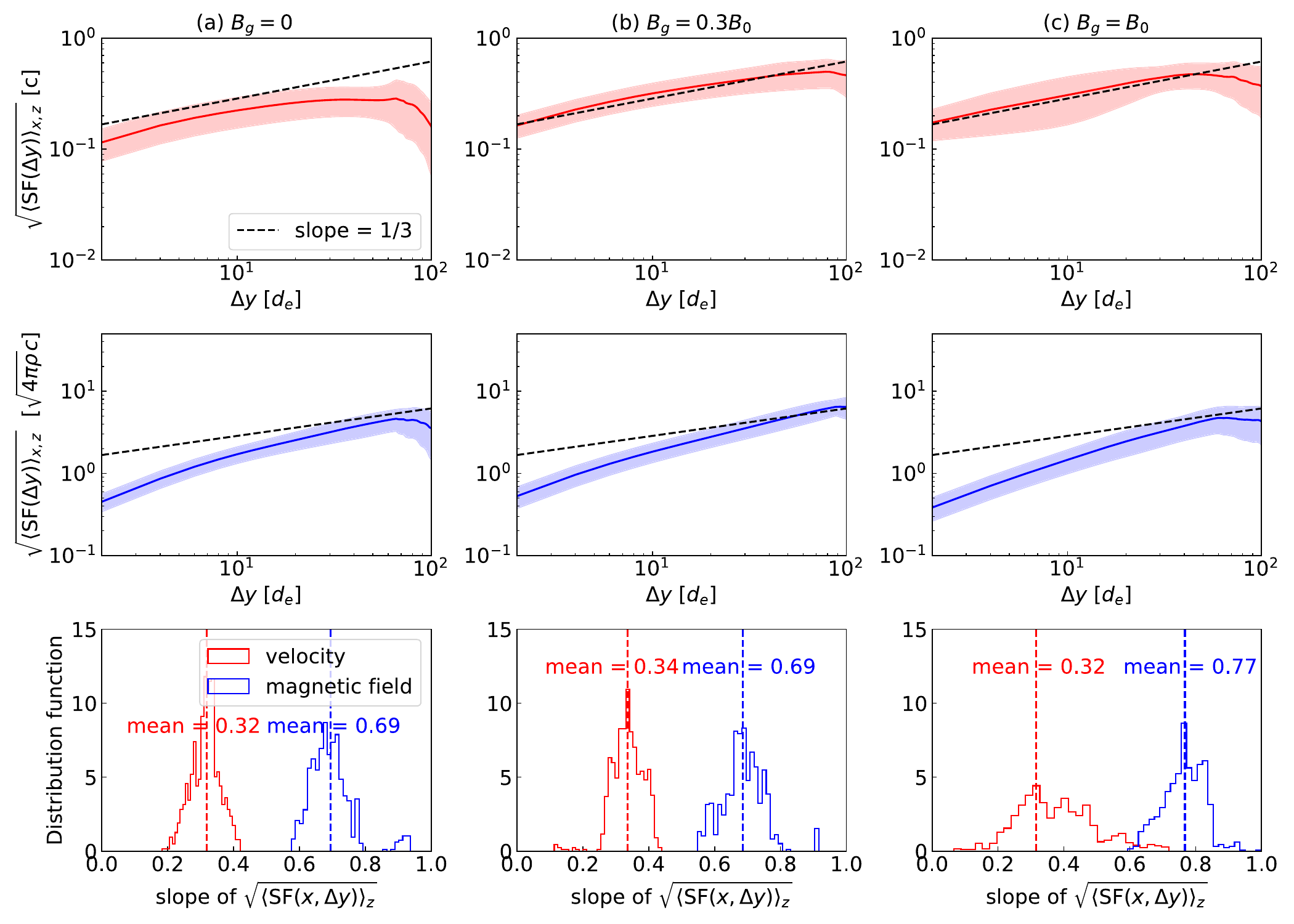}
        \caption{\textbf{Top and middle panels:} Square root of the second-order structure function (SF) for the velocity (red) and magnetic field (blue). The SF is computed for separations along the $y$-axis at each position $(x, z)$. The resulting ${\rm SF}(x,\Delta y,z)$ are then averaged over all $x$ and $z$ values, denoted as $\langle{\rm SF}(\Delta y)\rangle_{z,x}$. The shaded area represents the standard deviation calculated from the data points at a given spatial separation. The black dashed line indicates the slope of 1/3. Left panels show results without a guide magnetic field; the middle and right panels include guide fields $B_g=0.3B_0$ and $B_g=B_0$, respectively. \textbf{Bottom panel:} The distribution of the $\langle{\rm SF}(\Delta y, x)\rangle_z$'s slope. The slope is fitted over separations in the range 2 – 30$d_e$. The red and blue dashed lines represent the mean slopes of the velocity and magnetic field structure functions, respectively.}
    \label{fig:sf_dx}
\end{figure*}

\subsubsection{Outflow ($x$) direction}
Fig.~\ref{fig:sf_dy} presents the square root of the second-order structure function computed for separations along the $x$-axis at each $(y, z)$ location and subsequently averaged over all $y$ and $z$ within the reconnection region. For the velocity fluctuations, in the absence of a guide field, $\sqrt{\langle{\rm SF}_v(\Delta x)\rangle_{y,z}}$ follows an approximate power law between 2 and 30$d_e$, with a mean slope of $\sim$ 0.415, similar to the scaling measured for separations along the $z$–direction (see Fig.~\ref{fig:sf_xmid}). The slope increases at separations larger than $\sim300d_e$, which could be an effect of the net outflow.
When a strong guide field $B_g=B_0$ is included, the mean slope at scales below 30$d_e$ becomes very close to 1/3, while the amplitude of the structure function increases slightly compared to the case without a guide field at small scales. At larger scale separations ($\gtrsim$~30$d_e$), the velocity structure function flattens in all cases, more prominently so as the guide field increases. 

For the magnetic fluctuations, $\sqrt{\langle{\rm SF}_B(\Delta x)\rangle_{y,z}}$ is again systematically steeper than the square root of the velocity structure function. The mean slopes are $\sim$ 0.665 without a guide field, and $\sim$0.715 and 0.658 for $B_g=0.3B_0$ and $B_g=B_0$, respectively. In the case of vanishing guide field, the slope is also larger than that of $\sqrt{\langle{\rm SF}_B(\Delta z)\rangle_{(x,y_{\rm mid})}}$. This enhanced steepness along the outflow direction may reflect the influence of reconnection outflows, which tend to generate strong magnetic gradients at large scales. A finite guide field increases the variability of the slopes in the outflow direction, widening their overall distribution. We caution that the fitted scaling exponents are extracted over a relatively limited range of scales, set by the competing requirements of resolving the plasma skin depth and capturing system-scale dynamics. Future simulations with larger scale separation between the system size and $d_e$ will be important to assess whether the reported scalings represent a well-developed inertial range.

\subsubsection{Inflow ($y$) direction}
Fig.~\ref{fig:sf_dx} presents the square root of the second-order structure function computed for separations along the inflow ($y$) direction at each $(x, z)$ location and subsequently averaged over all $x$ and $z$. For the velocity fluctuations, in the absence of a guide field, $\sqrt{\langle{\rm SF}_v(\Delta y)\rangle_{x,z}}$ generally exhibits a power-law scaling over $\sim4 - 20d_e$, though it appears mildly curved, likely due to some memory of the net inflow motion. The power-law scaling is clearer in the presence of a guide field and extends to $\sim 40d_e$. The mean slopes are approximately 0.32 without a guide field, and 0.34 and 0.32 for $B_g = 0.3B_0$ and $B_g =B_0$, respectively.

For the magnetic fluctuations, $\sqrt{\langle{\rm SF}_B(\Delta y)\rangle_{x,z}}$ displays mean slopes of $\sim0.69$ in both the $B_g = 0$ and $0.3B_0$ cases, comparable to those along the outflow ($x$) and out-of-plane ($z$) directions. The slope increases to 0.77 when $B_g = B_0$. The presence of a strong guide field also enhances the spread of both velocity and magnetic structure-function slopes, broadening their distributions.


\subsubsection{Dependence on the guide field strength}
\begin{figure*}
\centering
\includegraphics[width=0.99\linewidth]{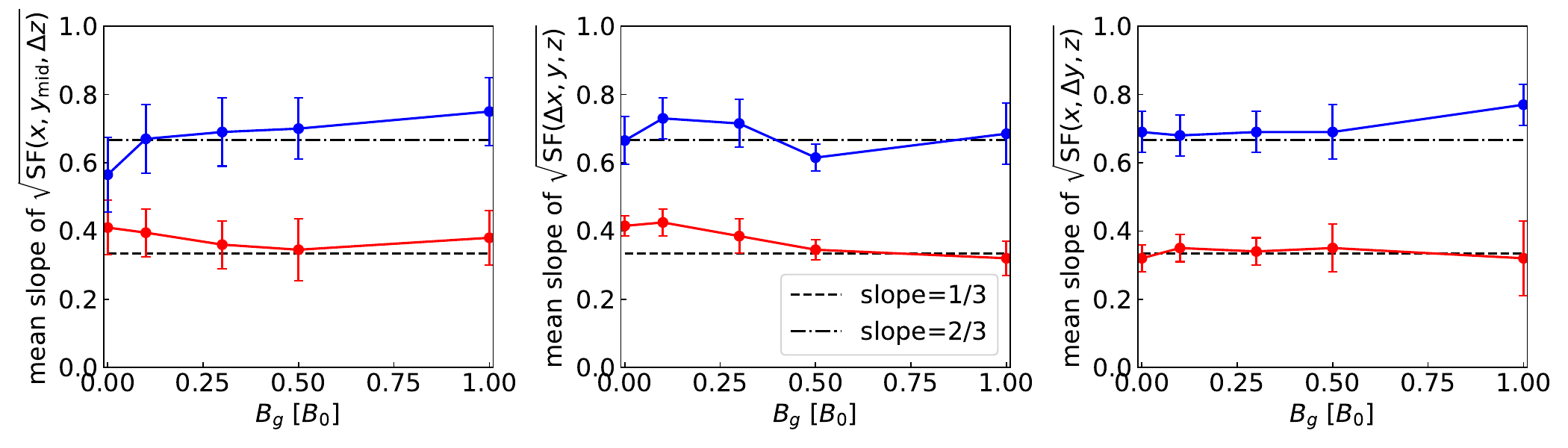}
        \caption{Mean slopes of $\sqrt{{\rm SF}({x,y_{\rm mid}, \Delta z)}}$ (left), $\sqrt{{\rm SF}(\Delta x,y,z)}$ (middle), $\sqrt{{\rm SF}( x,\Delta y,z)}$ (right) as a function of the guide field strength. The slope is fitted over separations in the range 2 – 30$\, d_e$. The black dashed line indicates the slope of 1/3, while the black dotted-dashed line indicates the slope $2/3$. The error bars show the one sigma standard deviations of the fitted parameters (estimated from the covariance matrix of the least-squares fit).}
    \label{fig:sf_slope}
\end{figure*}

In Fig.~\ref{fig:sf_slope}, we summarize how the mean slopes of $\sqrt{{\rm SF}(x, y_{\rm mid}, \Delta z)}$, $\sqrt{{\rm SF}(\Delta x, y, z)}$, and $\sqrt{{\rm SF}(x, \Delta y, z)}$ vary with guide-field strength. For the velocity fluctuations, the guide field has little influence, with the slopes remaining close to $1/3$, ranging from $\sim0.3$ to $\sim0.4$. A weak trend is nevertheless apparent, in which the mean slope of $\sqrt{{\rm SF}_v(\Delta x, y, z)}$ slightly decreases as the guide field increases. In contrast, the magnetic fluctuations exhibit progressively steeper scaling with increasing guide-field strength, most clearly in the mean slopes of $\sqrt{{\rm SF}B(x, y{\rm mid}, \Delta z)}$ and $\sqrt{{\rm SF}_B(x, \Delta y, z)}$. Overall, the square-root second-order magnetic structure-function slopes are substantially steeper than $1/3$, typically clustering around $2/3$, with values ranging from $\sim0.6$ to $\sim0.8$ across the cases examined.

In Appendix Figs.~\ref{fig:sf_slope_005_th10} and \ref{fig:sf_slope_005_th20}, we repeat the analysis using density thresholds of 10\% and 20\%, respectively, to define the reconnection regions. For both the thresholds, the values and trends of the velocity and magnetic-field structure-function slopes as functions of guide-field strength are similar to those in Fig.~\ref{fig:sf_slope}, which adopts the 15\% threshold.

In Appendix Fig.~\ref{fig:sf_slope_005_back}, we isolate the turbulent fluctuations from the large-scale reconnection flow and field by removing the background components. Specifically, before calculating the structure functions, we subtract the $z$-averaged profile from each velocity and magnetic-field component. This procedure removes the reconnection exhaust and inflow profiles, as well as any residual large-scale magnetic structure associated with the mean (i.e., $z$-averaged)) reconnection geometry.

Fig.~\ref{fig:sf_slope_005_back} shows the mean slopes of $\sqrt{{\rm SF}(x, y_{\rm mid}, \Delta z)}$, $\sqrt{{\rm SF}(\Delta x, y, z)}$, and $\sqrt{{\rm SF}(x, \Delta y, z)}$ as functions of guide-field strength using the mean-subtracted velocity and magnetic fields. Compared with Fig.~\ref{fig:sf_slope}, the mean slopes of $\sqrt{{\rm SF}(x, y_{\rm mid}, \Delta z)}$ for both the velocity and magnetic fields increase slightly, by about 0.1, when the guide field exceeds $0.5B_0$. Meanwhile, the slopes of $\sqrt{{\rm SF}_B(x, \Delta y, z)}$ decrease by $\sim0.1$--$0.2$, largely independent of guide-field strength. The remaining slopes are consistent with those obtained using the unsubtracted velocity and magnetic fields.

\begin{figure*}
\centering
\includegraphics[width=0.99\linewidth]{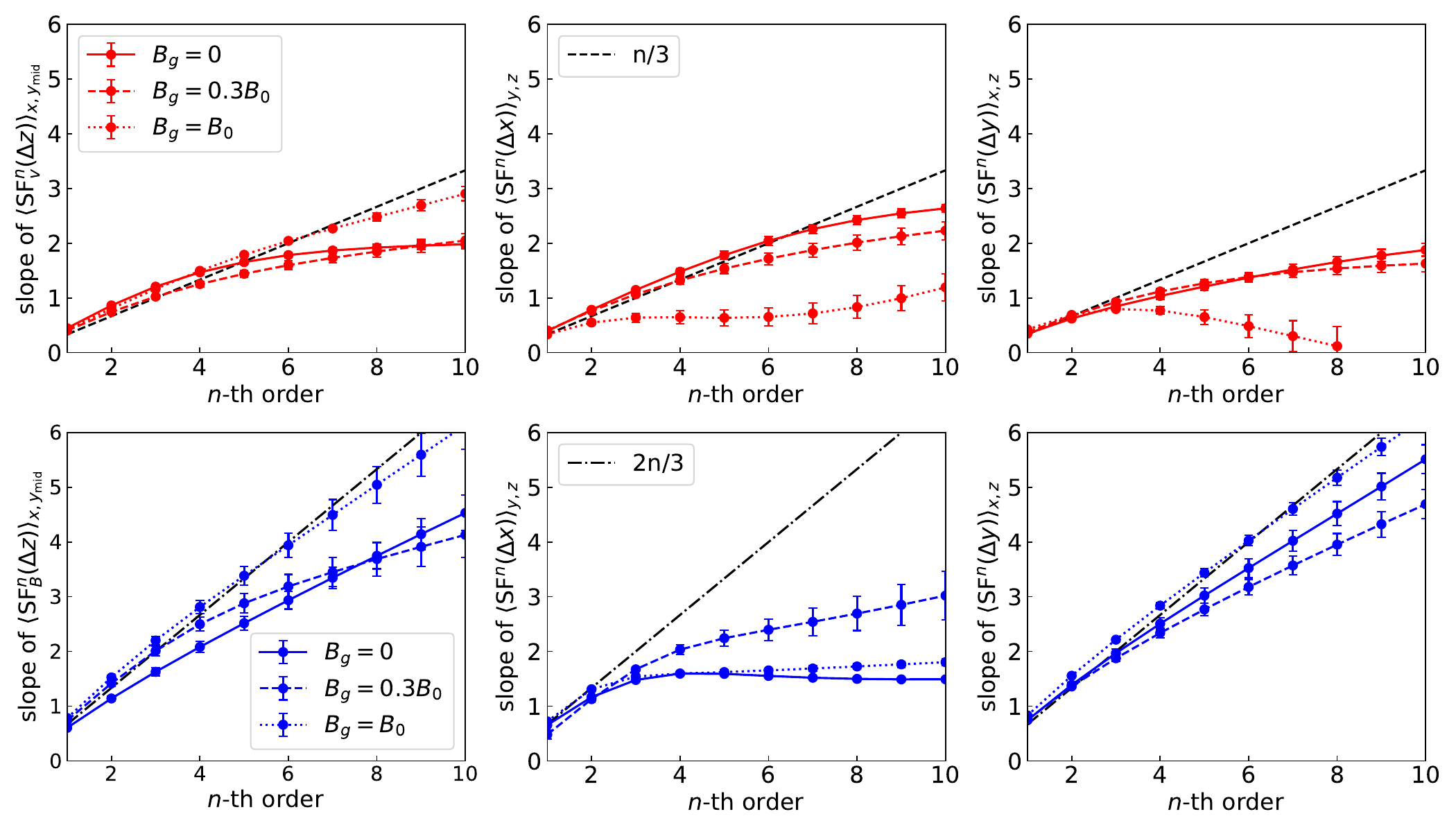}
        \caption{Slopes of the $n$-th order structure functions: $\langle{\rm SF}^n(\Delta z)\rangle_{x,y_{\rm mid}}$ (left), $\langle{\rm SF}^n(\Delta y)\rangle_{x,z}$ (middle), $\langle{\rm SF}^n(\Delta x)\rangle_{y,z}$ (right). The slope is fitted over separations in the range 2 – 30$\, d_e$. Intermittency is reflected in departures from linear, self-similar scaling. The black dashed line indicates the expected $n/3$ for self-similar Kolmogorov-type fluctuations without intermittency, while the black dotted-dashed line indicates the slope $2n/3$. The error bars show the one sigma standard deviations of the fitted parameters (estimated from the covariance matrix of the least-squares fit).}
    \label{fig:sf_nth}
\end{figure*}

\subsection{High‑order structure functions and intermittency}

To quantify the intermittency of the reconnection-driven turbulent fluctuations, we compute the $n$‑th order structure functions along the guide field ($z$), inflow ($y$), and outflow ($x$) directions. As an example, the $n$-th order structure function along the guide field direction is defined as:
\begin{equation}
{\rm SF}^{n}(x,y,\Delta z)=\langle|\pmb{f}(x,y,z)-\pmb{f}(x,y,z+\Delta z)|^n\rangle,
\end{equation}
where $\pmb{f}$ denotes either the velocity or the magnetic field. In the structure-function formalism, intermittency appears as a departure from the linear, self-similar scaling of the structure-function exponents $n/\xi$, where $\xi$ is a constant number determined by self-similar behaviors. For instance, a self-similar Kolmogorov scenario predicts a slope of $n/3$ (black dashed line in Fig.~\ref{fig:sf_nth}). For non-Kolmogorov type similarity, $\xi$ could be different. Greater deviations from linear scaling, particularly the flattening at large $n$, are a signature of stronger intermittency, as high-order exponents become increasingly dominated by rare, intense structures. 

Fig.~\ref{fig:sf_nth} presents the scaling exponents, or slopes, of the velocity and magnetic-field structure functions for orders $n=1$--$10$. For the velocity fluctuations, the exponents approximately follow the linear Kolmogorov prediction, $n/3$, at low orders ($n<5$) along the guide-field ($z$) direction for all guide-field strengths, but deviate substantially at higher orders. This behavior reflects the non-Gaussian nature of the fluctuations, as also shown by the PDFs in Appendix Figs.~\ref{fig:pdf_l_005} and \ref{fig:pdf_b_005}. Along the inflow ($y$) direction, the curves flatten more noticeably already for $n>3$ compared with the $z$ direction. The departure from linear self-similarity is most pronounced along the outflow ($x$) and inflow ($y$) directions for the $B_g=B_0$ case.

It is important to note, however, that these exponents are globally averaged quantities. Locally, the scaling can vary substantially; for example, the second-order structure-function slope shows a dispersion of $\sim0.2$--$0.3$ (see Figs.~\ref{fig:sf_xmid}, \ref{fig:sf_dy}, and \ref{fig:sf_dx}). Standard intermittency models are typically formulated for a well-defined inertial cascade and implicitly assume a large separation between forcing and dissipation scales \citep{1994PhRvL..72..336S,1997PhFl....9.3817G,2025NPGeo..32..243P}. In our reconnection regions, the turbulence is strongly inhomogeneous, and energy injection may occur over a range of scales. Under these conditions, a quantitative comparison with a specific intermittency model would not be well posed. We therefore present the high-order exponents here only as descriptive trends.

In addition, accurately resolving high-order exponents requires high numerical resolution. Ideally, dedicated convergence tests in particles per cell (ppc) and spatial resolution (i.e., cells per skin depth) \citep{2021ApJ...922..261Z}, together with an ensemble of simulations using different random seeds, should be performed to validate these results directly, especially the high-order intermittency exponents. Such tests are computationally challenging, however, given the large three-dimensional domain sizes required. Whether the high-order exponents are quantitatively sensitive to these numerical choices remains an important question for future PIC studies.

Regarding magnetic field fluctuations, Fig.~\ref{fig:sf_nth} indicates that for all guide-field strengths (including $B_g = 0$ and $B_g=B_0$), the scaling exponents display pronounced flattening along the outflow ($x$) direction. This flattening is more significant than that observed in the velocity field, suggesting that stronger intermittency occurs in the magnetic field. In contrast, regardless of the guide-field strength, the scaling exponents increase nearly linearly with $n$ along the guide-field ($z$) and inflow ($y$) directions, with only moderate flattening occurring at high orders ($n>5$). Specifically, the flattening along the inflow ($y$) direction is less prominent than in the velocity field (indicating weaker intermittency), while the behavior in the guide-field ($z$) direction is comparable to that of the velocity field.

\begin{figure*}
\centering
\includegraphics[width=0.99\linewidth]{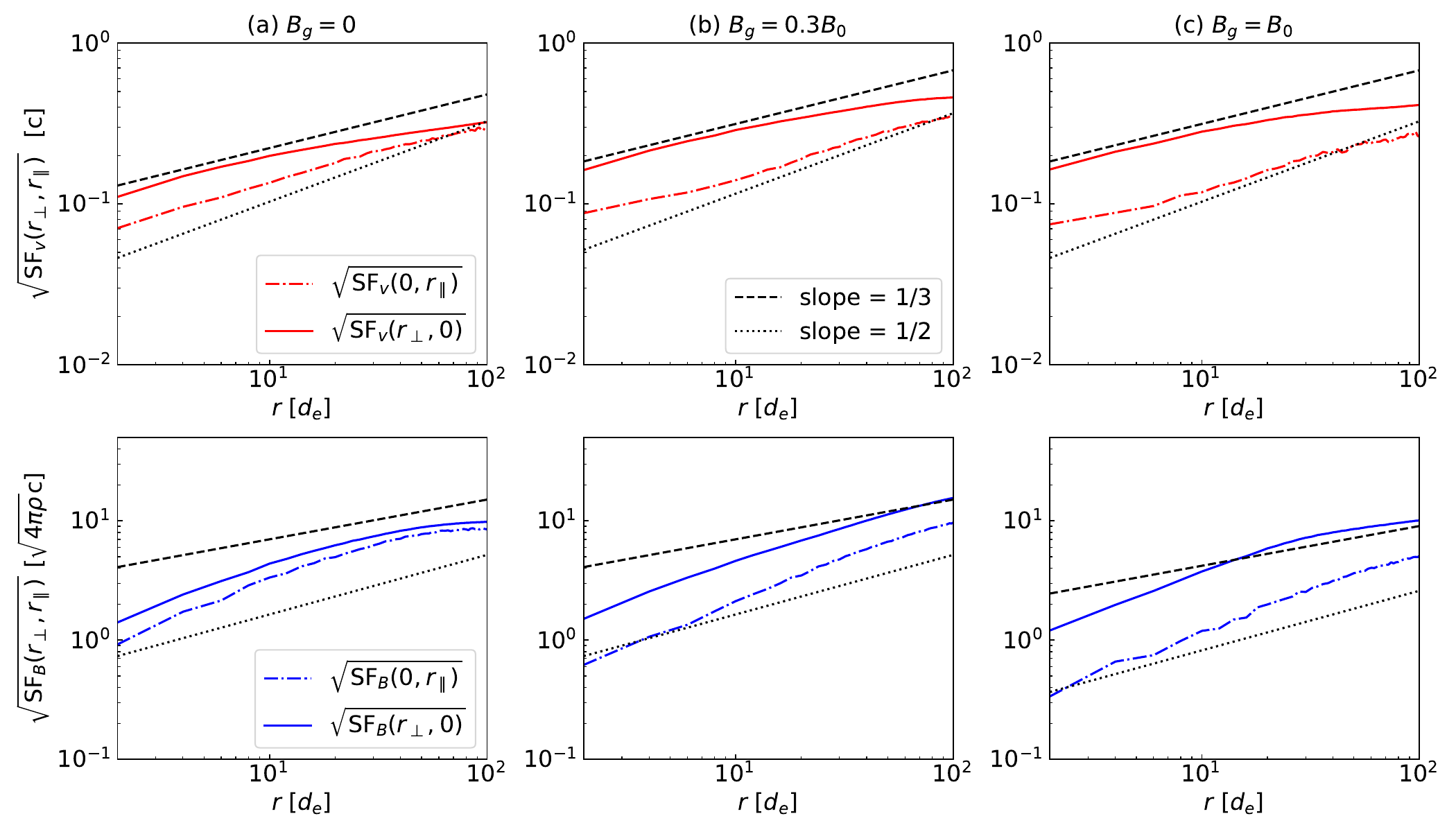}
        \caption{Square root of the second-order structure function (SF) for the velocity (red) and magnetic field (blue) decomposed into components parallel and perpendicular to the local magnetic field. The black dashed and dotted lines indicate the scaling slopes of 1/3 and 1/2, respectively. Left panels show results without a guide magnetic field; middle and right panels include guide fields $B_g=0.3B_0$ and $B_g=B_0$, respectively.}
    \label{fig:sf_anisotropy}
\end{figure*}
\subsection{Anisotropy of velocity and magnetic field fluctuations}
To characterize anisotropy in the fluctuations within the reconnection layer, we decompose the structure functions into components parallel and perpendicular to the local magnetic field. This procedure follows the methodology developed by \citet{CV20}, averaging over all directions and locations within the reconnection region\footnote{Due to the limited number of data points along any single axis, we did not compute structure functions specifically along the $x$, $y$, or $z$ directions in this subsection.}:
\begin{equation}
\label{eq.sf_loc}
\begin{aligned}
&\pmb{B}=\frac{1}{2}(\boldsymbol{B}(\pmb{x}+\pmb{r})+\pmb{B}(\pmb{x})),\\
&{\rm SF}(r_\bot,r_\parallel)=\langle|\pmb{f}(\pmb{x}+\pmb{r})-\pmb{f}(\pmb{x})|^2\rangle_{\pmb{x}},
\end{aligned}
\end{equation}
where $\pmb{B}$ defines the local magnetic-field direction in a local cylindrical coordinate system, with $\hat{r}_\parallel=\pmb{B}/|\pmb{B}|$, $r_\bot=|\hat{r}_\parallel\times\pmb{r}|$, and $r_\parallel=\hat{r}_\parallel\cdot\pmb{r}$. From these, we derive the parallel and perpendicular fluctuations relative to the local magnetic field, ${\rm SF}(0,r_\parallel)$ and ${\rm SF}(r_\bot,0)$, respectively. 

Fig.~\ref{fig:sf_anisotropy} presents the square root of these structure functions for both velocity and magnetic field fluctuations, comparing the cases with guide field $B_g=0$, $B_g=0.3B_0$, and $B_g=B_0$. In all cases, clear anisotropies are present, with perpendicular fluctuations dominating over parallel ones. For the velocity fluctuations, the perpendicular component $\sqrt{SF_v(r_\perp,0)}$ follows a slope close to $1/3$ up to 20$d_e$, whereas the parallel component $\sqrt{SF_v(0,r_\parallel)}$ exhibits a slope in between 1/3 and 1/2. Importantly, the anisotropy increases at small scales. This indicates that the velocity fluctuation anisotropy—specifically, the ratio of perpendicular to parallel components—is scale-dependent, becoming more pronounced at smaller scales. This behavior is consistent with theoretical expectations for magnetized turbulence \citep{LV99,CV20,2021ApJ...911...37H}. The presence of a guide field further amplifies this anisotropy, yielding a larger ratio of perpendicular to parallel fluctuations.

In the absence of a guide field, magnetic-field fluctuations exhibit weaker anisotropy than velocity fluctuations, but with distinct trends. Both parallel and perpendicular magnetic structure functions scale more steeply than $1/2$, with the perpendicular component remaining somewhat larger than the parallel one. With a moderate guide field, $B_g=0.3B_0$, both components steepen further, and the difference between $\sqrt{{\rm SF}_B(r_\perp,0)}$ and $\sqrt{{\rm SF}_B(0,r_\parallel)}$ widens. The difference between $\sqrt{{\rm SF}_B(r_\perp,0)}$ and $\sqrt{{\rm SF}_B(0,r_\parallel)}$ further increases with a stronger guide field, $B_g=B_0$, and the scaling slopes remain steeper than $1/2$, with values closer to 2/3. Additionally, we tested the robustness of our results by excluding points within the reconnection regions where the magnetic field strength falls in the lowest 3rd percentile. We found that the results remain consistent with this exclusion.

\section{Discussion}
The structure-function analysis presented above provides a statistical characterization of the fluctuations self-consistently generated by collisionless magnetic reconnection in the magnetically dominated regime. In a reconnection layer, kinetic energy is injected across a range of scales through flux-rope dynamics and reconnection-driven outflows, and coherent structures coexist with the developing turbulent fluctuations. This environment differs from the idealized setting assumed in classical turbulence phenomenology, where a wide separation between forcing and dissipation scales is implicit. The measured structure-function exponents therefore reflect the aggregate scaling behavior of reconnection-driven fluctuations in this inhomogeneous, multiscale system, rather than necessarily isolating a strictly universal inertial-range cascade. The same consideration applies to the high-order intermittency analysis, for which we have already noted that a quantitative comparison to standard intermittency models is not well-posed.

Regarding numerical resolution, the PIC simulations employ 2 particles per cell (ppc) in the upstream region. In the relativistic, magnetically dominated regime, the plasma dynamics is primarily governed by the electromagnetic fields rather than thermal particle pressure, reducing the sensitivity of macroscopic statistics to particle noise compared to non-relativistic or weakly magnetized cases. In addition, the reconnection downstream is generally overdense as compared to the upstream, implying effectively a higher number of particles per cell in the regions considered for our analysis. We therefore expect our choice of ppc to be sufficient for the second order structure-function statistics reported here. 

Within this framework, the key results of this study are robust and physically meaningful. The velocity fluctuations exhibit slopes broadly consistent with 1/3, while magnetic fluctuations are systematically steeper around 2/3; both fields exhibit scale-dependent anisotropy with respect to the local magnetic field, and intermittency is strongest along the reconnection outflow direction. These are consistent features of the reconnection-driven fluctuations across guide-field strengths and simulation times. The 1/3 velocity scaling is particularly noteworthy, emerging self-consistently from a collisionless, magnetically dominated plasma without any external turbulent forcing.

\section{Conclusion} 
\label{sec:conclusion}
We have analyzed a suite of large-scale 3D PIC simulations to investigate the statistical properties of velocity and magnetic-field fluctuations generated by collisionless magnetic reconnection in a magnetically-dominated plasma. Our analysis examined how the scaling, anisotropy, and intermittency of these fluctuations depend on the strength of a background guide field, spanning the values $B_g = 0, 0.1B_0, 0.3B_0, 0.5B_0,$ and $B_0$.

Our first-principles PIC simulations show that magnetic reconnection self-consistently drives multiscale velocity and magnetic fluctuations, extending from system-size scales down to scales comparable to the plasma skin depth. The square root of the second-order structure functions of the velocity fluctuations exhibits a $1/3$ scaling, whereas magnetic fluctuations display systematically steeper slopes, typically near $2/3$, but ranging from $\sim0.6$ to $\sim0.8$ across the guide-field strengths examined. Future work should explore whether alternative diagnostics can independently corroborate the scaling results reported here.

The morphology of the fluctuations depends on the guide field. In the absence of a guide field, velocity and magnetic fluctuations are strongly elongated along the outflow direction. Introducing a guide field progressively reorients these structures to align with the guide field direction. The guide field also reduces the overall amplitude of magnetic fluctuations and broadens the distribution of scaling slopes for both velocity and magnetic structure functions. While a strong guide field slightly steepens the magnetic-field structure function slope, its effects on the velocity structure function slope are not apparent.

High-order structure functions reveal that intermittency in magnetic fluctuations has a directional difference. The strongest magnetic intermittency occurs along the outflow ($x$) direction, exceeding that of the velocity field. The inflow ($y$) direction exhibits weaker intermittency than the velocity field, while the guide-field ($z$) direction is comparable. However, further ppc-convergence and random-seed tests may be needed for confirming the high-order intermittency exponents.

Anisotropy analysis, based on structure functions decomposed parallel and perpendicular to the local magnetic field, shows that both velocity and magnetic fluctuations are more pronounced across the field than along it. This scale-dependent anisotropy is stronger in the velocity fluctuations than in the magnetic fluctuations for small or moderate guide fields. The presence of a guide field enhances the anisotropy in both the velocity and magnetic fluctuations. 

Overall, these results highlight how reconnection not only releases magnetic energy but also generates turbulent fluctuations with distinct scaling, anisotropy, and intermittency signatures. The sensitivity of these properties to the presence of a guide field underscores its central role in shaping the characteristics of reconnection-driven turbulent fluctuations. Especially, the magnetic field fluctuations deviate from the 1/3 scalings over the available range of scales. The statistical features identified here—such as power-law scaling, anisotropy, and intermittency—may carry important implications for studies of plasma heating, particle acceleration, and the observational signatures of reconnection in magnetically dominated astrophysical environments \citep{2025ARA&A..63..127S}.

\begin{acknowledgments}
Y.H. acknowledges the support for this work provided by NASA through the NASA Hubble Fellowship grant No. HST-HF2-51557.001 awarded by the Space Telescope Science Institute, which is operated by the Association of Universities for Research in Astronomy, Incorporated, under NASA contract NAS5-26555. ChatGPT is used for proofreading. This work used SDSC Expanse CPU and NCSA Delta CPU through allocations PHY230032, PHY230033, PHY230091, PHY230105, PHY230178, and PHY240183, from the Advanced Cyberinfrastructure Coordination Ecosystem: Services \& Support (ACCESS) program, which is supported by National Science Foundation grants \#2138259, \#2138286, \#2138307, \#2137603, and \#2138296. L.C. acknowledges support from NSF grant PHY-2308944, NASA ATP award 80NSSC22K0667, and NASA ATP award 80NSSC24K1230. L.S. acknowledges support from DoE Early Career Award DE-SC0023015, NASA ATP 80NSSC24K1238, NASA ATP 80NSSC24K1826, and NSF AST-2307202. This work was supported by a grant from the Simons Foundation (MP-SCMPS-00001470) to L.S. and facilitated by Multimessenger Plasma Physics Center (MPPC), grant NSF PHY-2206609 to L.S. The data, code, and analysis scripts underlying this article will be shared on reasonable request to the corresponding author.
\end{acknowledgments}

%

\vspace{5mm}

\software{Python3 \citep{10.5555/1593511}, ChatGPT \citep{gpt}}

\appendix
\section{Comparison of the structure functions at time $t= \, 4.5 L/c$ and $t= \, 6.75 L/c$}

\begin{figure*}
  \centering
  \gridline{
    \fig{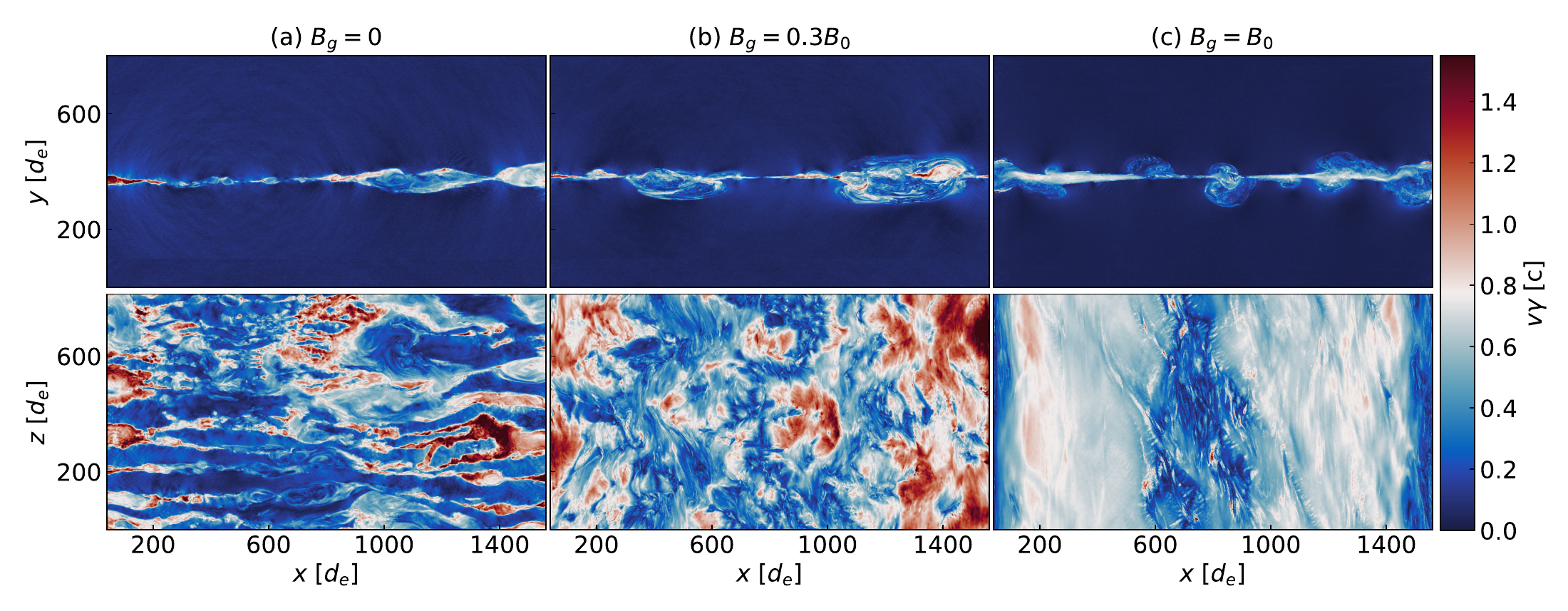}{1\textwidth}{}
  }
  \vspace{-4em}
  \gridline{
    \fig{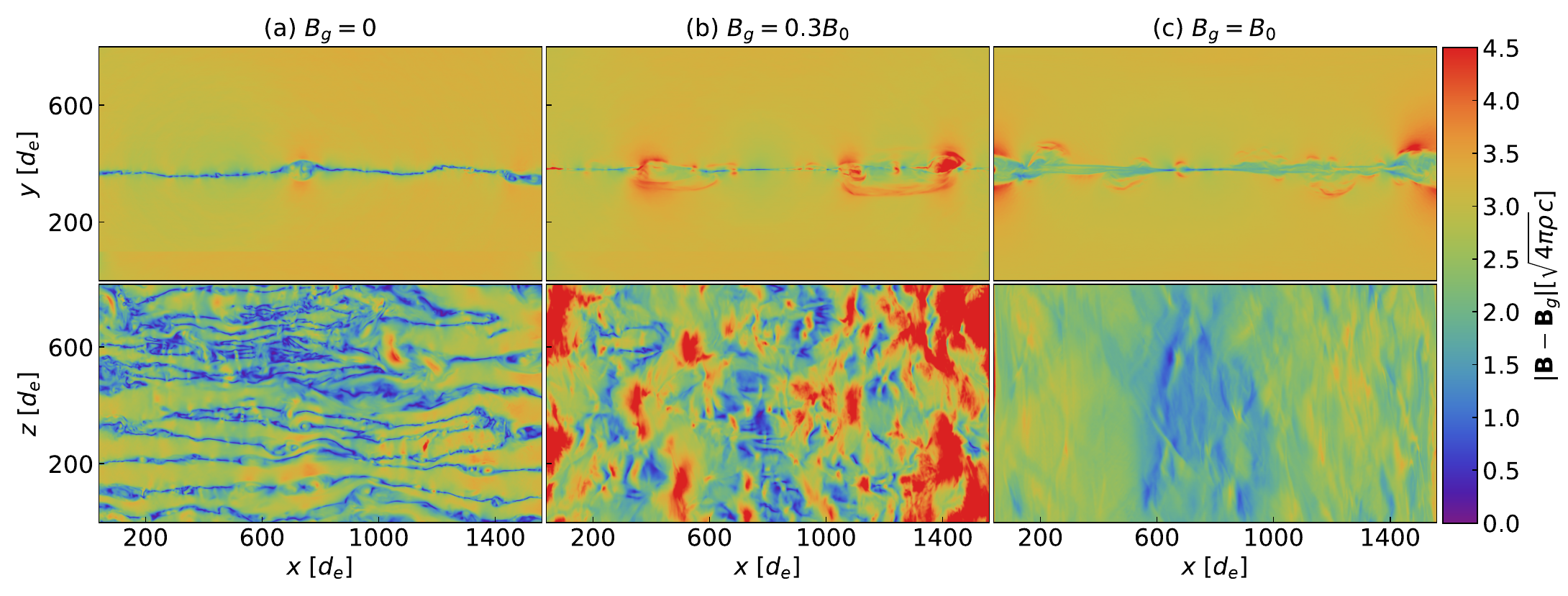}{1\textwidth}{}
  }
\vspace{-2.5em}
  \caption{Same as Fig.~\ref{fig:maps}, but for the time $t=4.5 \, L/c$.}
  \label{fig:maps_004}
\end{figure*}

\begin{figure*}
  \centering
  \gridline{
    \fig{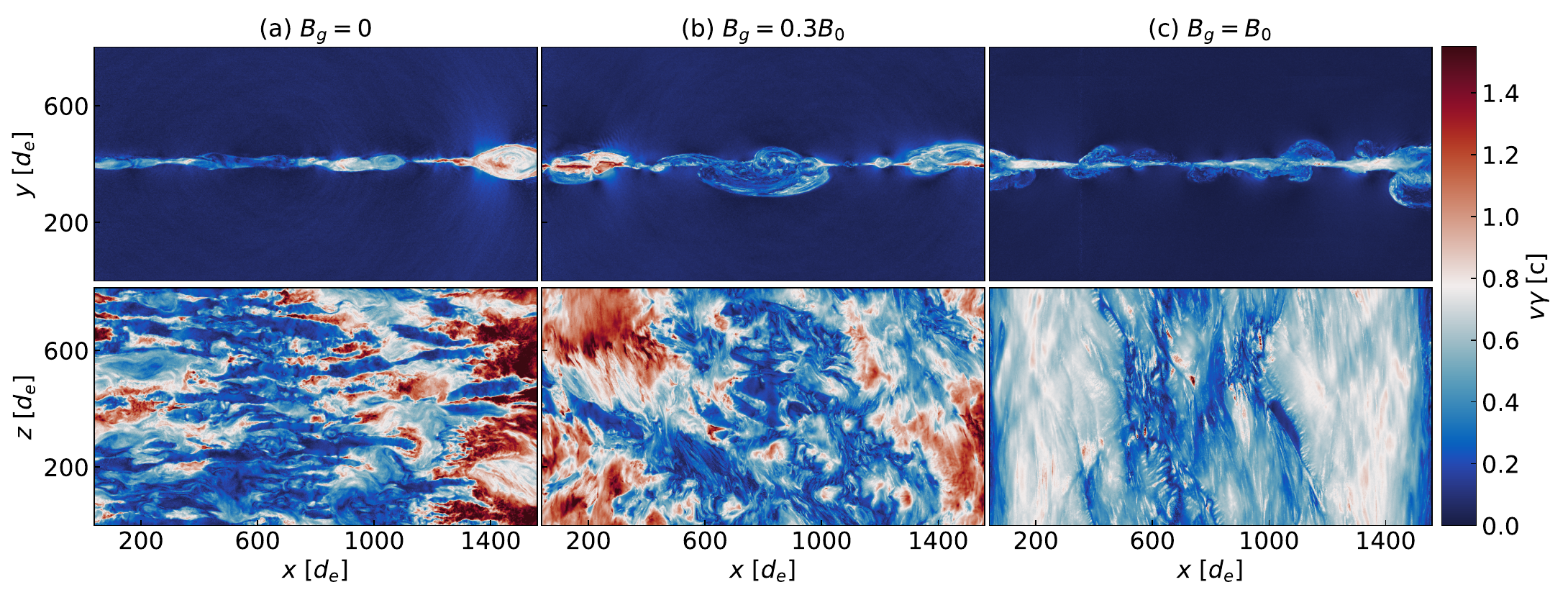}{1\textwidth}{}
  }
  \vspace{-4em}
  \gridline{
    \fig{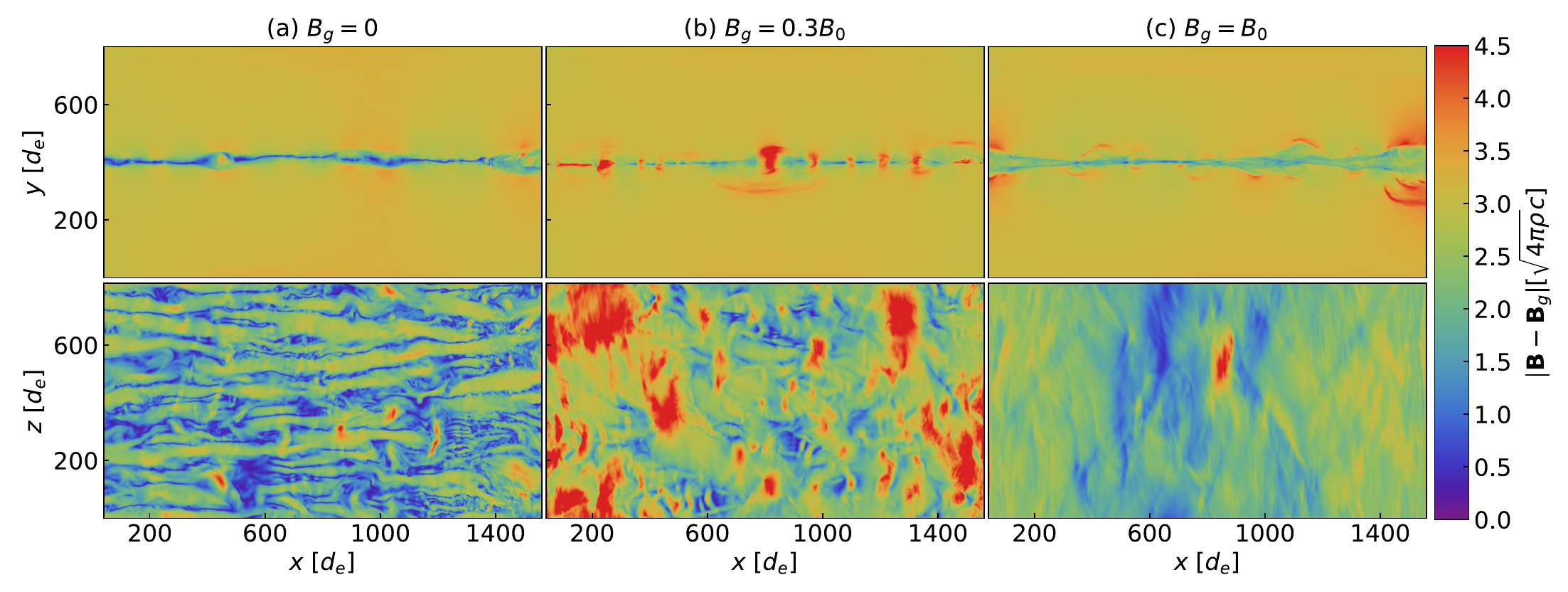}{1\textwidth}{}
  }
\vspace{-2.5em}
  \caption{Same as Fig.~\ref{fig:maps}, but for the time $t=6.75 \, L/c$.}
  \label{fig:maps_006}
\end{figure*}

Figs.~\ref{fig:maps_004} and \ref{fig:maps_006} show 2D slices of the velocity and magnetic fluctuations at $t=4.5\,L/c$ and $t=6.75\,L/c$, respectively, with guide-field strengths $B_g=0$, $0.3B_0$, and $B_0$. The $x-y$ slices are taken at $z$ location where the flux ropes are visually most prominent, while the $x-z$ slices are taken in the middle of the $y$ axis at the $z$ location.

\begin{figure*}
\centering
\includegraphics[width=0.5\linewidth]{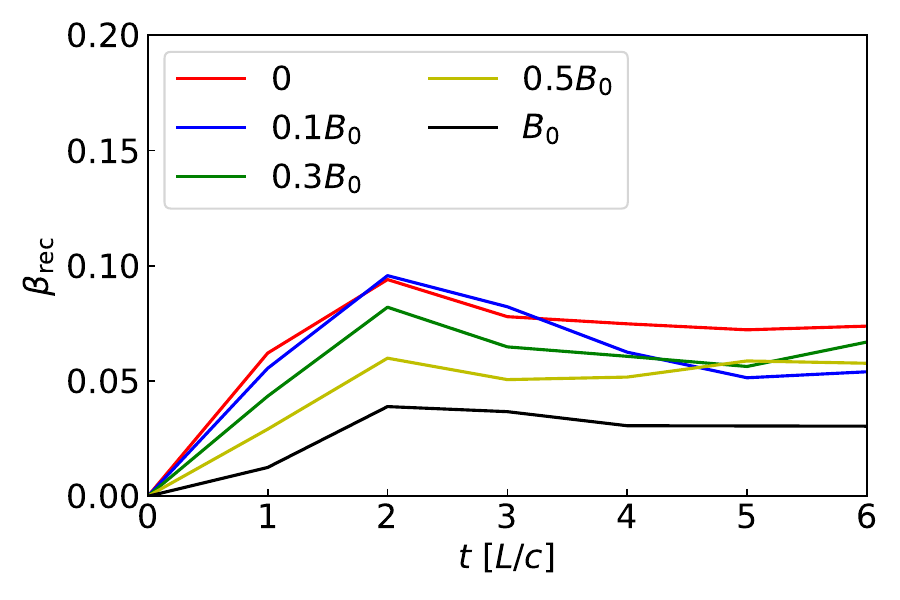}
        \caption{Time evolution of the reconnection rate $\beta_{\rm rec} = u_{\rm in}/c$, where $u_{\rm in}$ is the inflow speed for different strengths of the guide field, as indicated in the legend.}
    \label{fig:beta_rec}
\end{figure*}

\begin{figure*}
\centering
\includegraphics[width=0.99\linewidth]{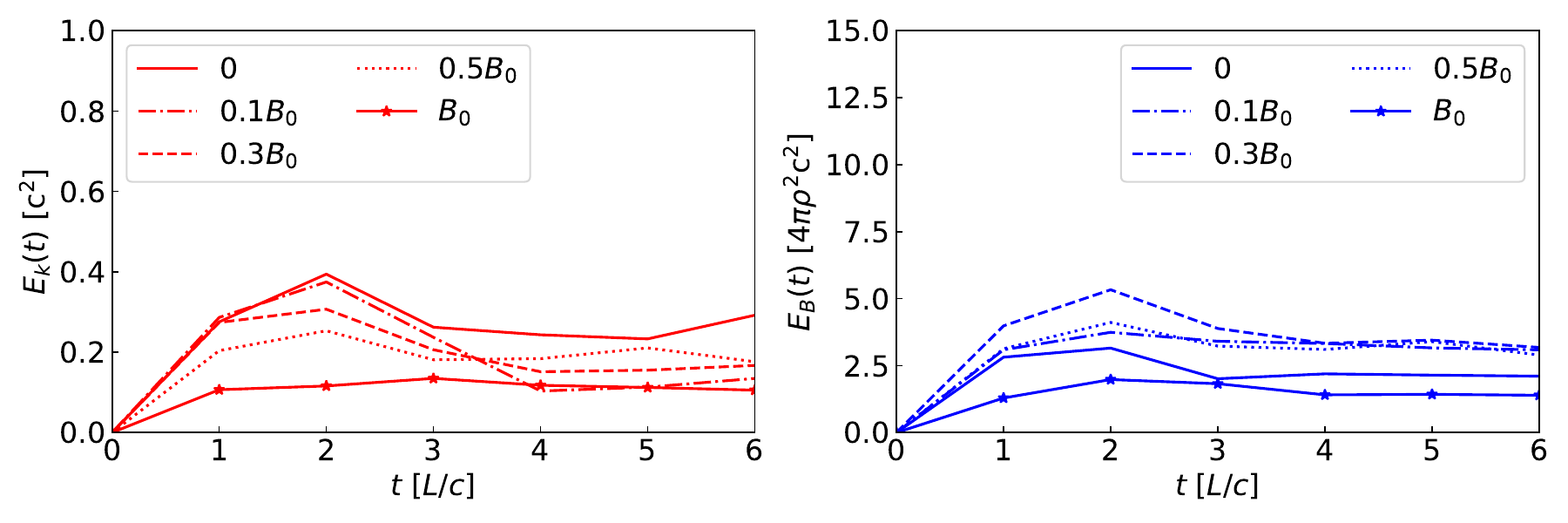}
        \caption{Time evolution of the kinetic energy and magnetic field fluctuation energy in the reconnection region.}
    \label{fig:EkEb}
\end{figure*}

Figs.~\ref{fig:beta_rec} and \ref{fig:EkEb} show the time evolution of the reconnection rate $\beta_{\rm rec} = u_{\rm in}/c$, kinetic energy, and magnetic field fluctuation energy. Here $u_{\rm in}$ is the inflow speed.

\begin{figure*}
\centering
\includegraphics[width=0.99\linewidth]{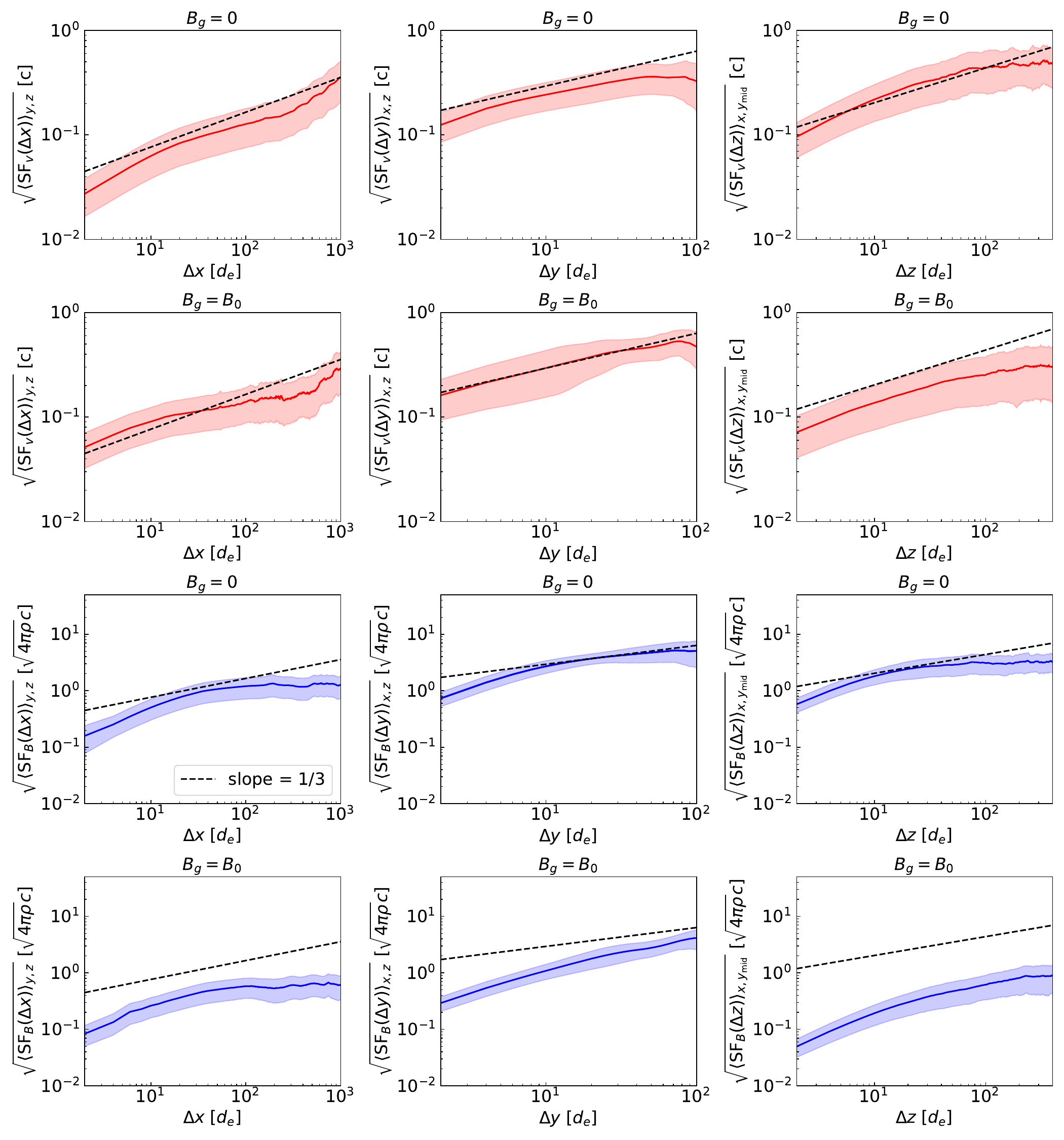}
        \caption{\textbf{First and second rows:} Square root of the second-order velocity structure functions, $\sqrt{\langle{\rm SF}_v(\Delta x)\rangle_{y,z}}$ (left), $\sqrt{\langle{\rm SF}_v(\Delta y)\rangle_{x,z}}$ (middle), and $\sqrt{\langle{\rm SF}_v(\Delta z)\rangle_{x,y_{\rm mid}}}$ (right), for $B_g=0$ and $B_g=B_0$. The shaded area represents the standard deviation calculated from the data points at a given spatial separation. The black dashed lines indicate the slope of 1/3. The structure functions are computed at time $t=4.5 \, L/c$. \textbf{Third and fourth rows:} Same as the first and second rows, but for the magnetic field structure functions.}
    \label{fig:sf_time_004}
\end{figure*}

\begin{figure*}[p]
\centering
\includegraphics[width=0.99\linewidth]{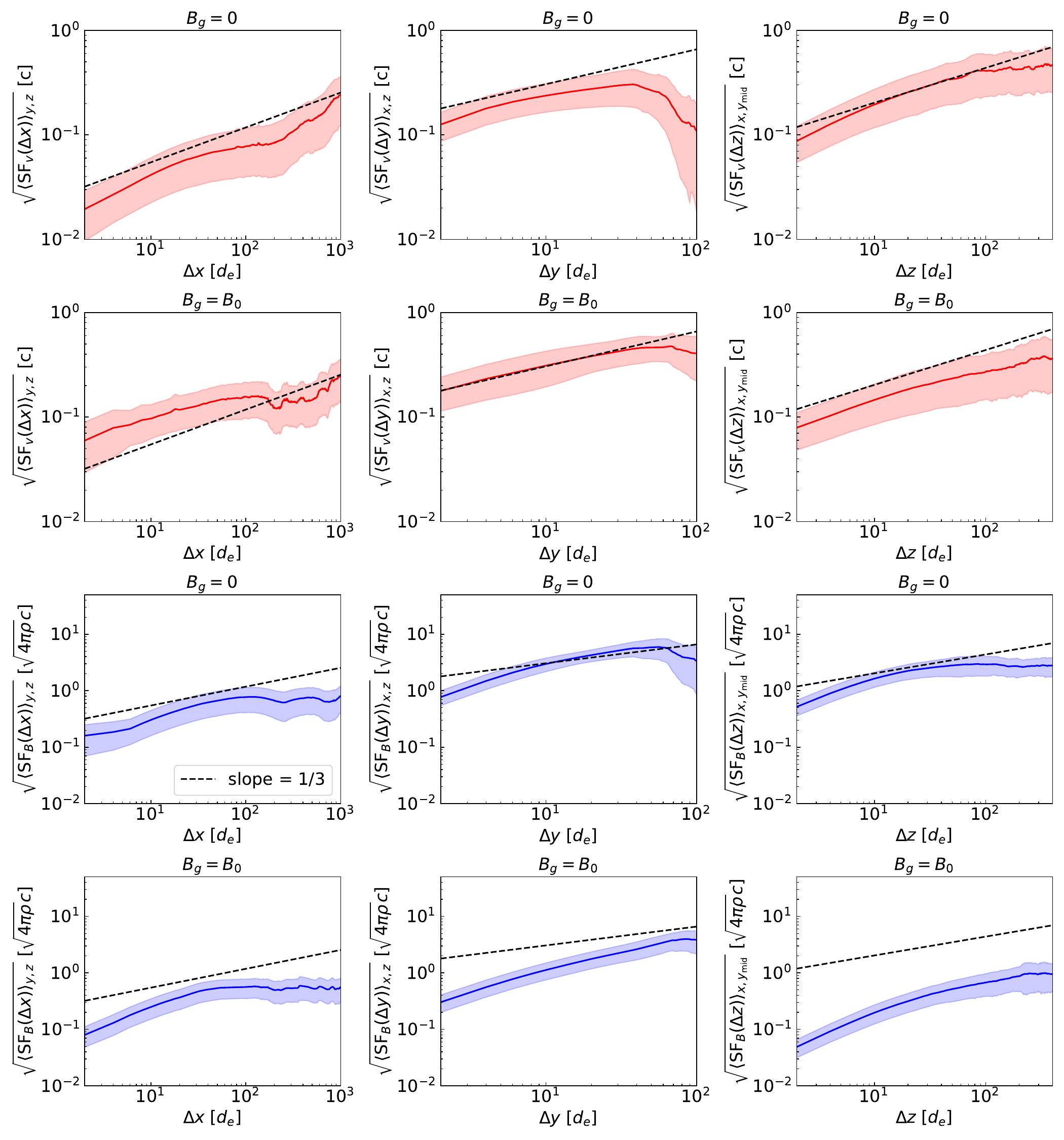}
        \caption{Same as Fig.~\ref{fig:sf_time_004}, but for the time $t=6.75 \, L/c$.}
    \label{fig:sf_time_006}
\end{figure*}

Figs.~\ref{fig:sf_time_004} and \ref{fig:sf_time_006} show the square root of the second-order structure functions $\sqrt{\langle{\rm SF}(\Delta x)\rangle_{y,z}}$ (left panels), $\sqrt{\langle{\rm SF}(\Delta y)\rangle_{x,z}}$ (middle panels), and $\sqrt{\langle{\rm SF}(\Delta z)\rangle_{x,y_{\rm mid}}}$ (right panels), computed at $t=4.5\,L/c$ and $t=6.75\,L/c$, respectively, for velocity (top two rows) and magnetic (bottom two rows) fluctuations. We present results for the limiting guide-field cases: $B_g=0$ and $B_g=B_0$. The variation of the slopes as a function of guide-field strength is detailed in Fig.~\ref{fig:sf_slope_006}. These structure functions exhibit behavior consistent with the main text and Figs.~\ref{fig:sf_xmid}--\ref{fig:sf_slope}: the velocity scaling slope is approximately $1/3$ up to $30d_e$, whereas the magnetic field structure function displays a steeper slope ($>1/3$).

\begin{figure*}
\centering
\includegraphics[width=0.99\linewidth]{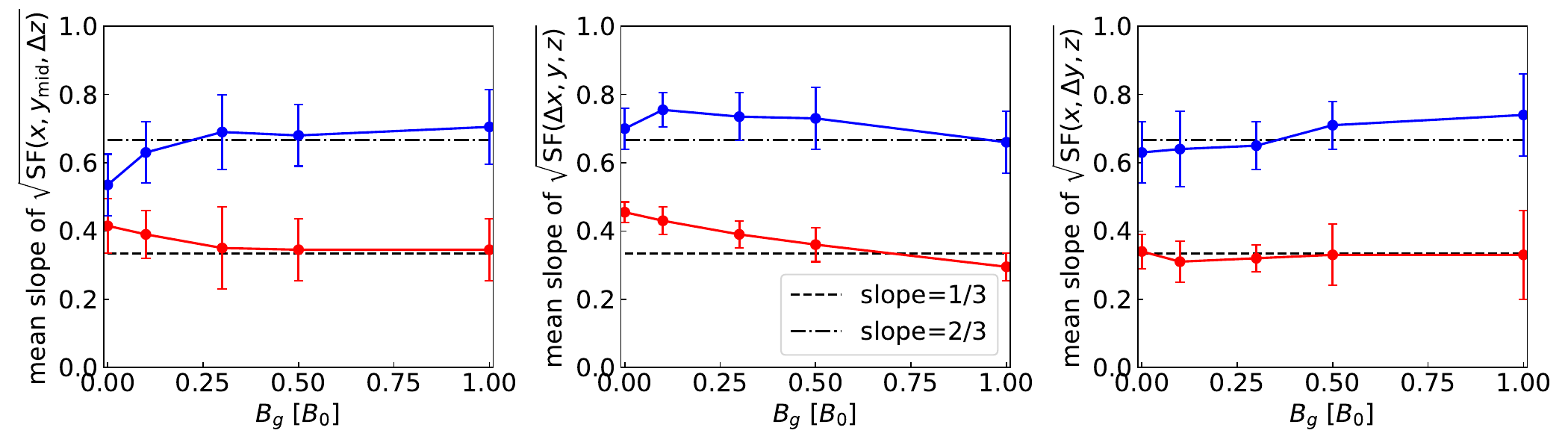}
        \caption{Mean slopes of $\sqrt{{\rm SF}({x,y_{\rm mid}, \Delta z)}}$ (left), $\sqrt{{\rm SF}(\Delta x,y,z)}$ (middle), $\sqrt{{\rm SF}( x,\Delta y,z)}$ (right) as a function of the guide field strength. The slope is fitted over separations in the range 2–30$\, d_e$. The black dashed line indicates the slope of 1/3, while the black dotted-dashed line indicates the slope $2/3$. The error bars show the one sigma standard deviations of the fitted parameters (estimated from the covariance matrix of the least-squares fit). The structure functions are computed at time $t=4.5 \, L/c$.} 
    \label{fig:sf_slope_004}
\end{figure*}

\begin{figure*}
\centering
\includegraphics[width=0.99\linewidth]{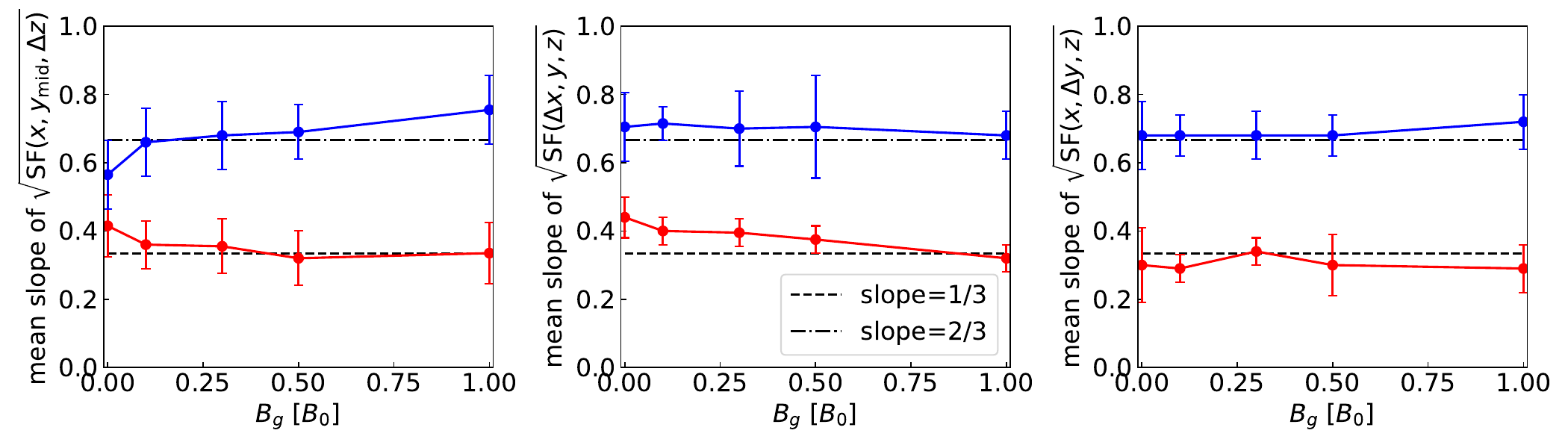}
        \caption{Same as Fig.~\ref{fig:sf_slope_004}, but for the time $t=6.75 \, L/c$.} 
    \label{fig:sf_slope_006}
\end{figure*}

\begin{figure*}
\centering
\includegraphics[width=0.99\linewidth]{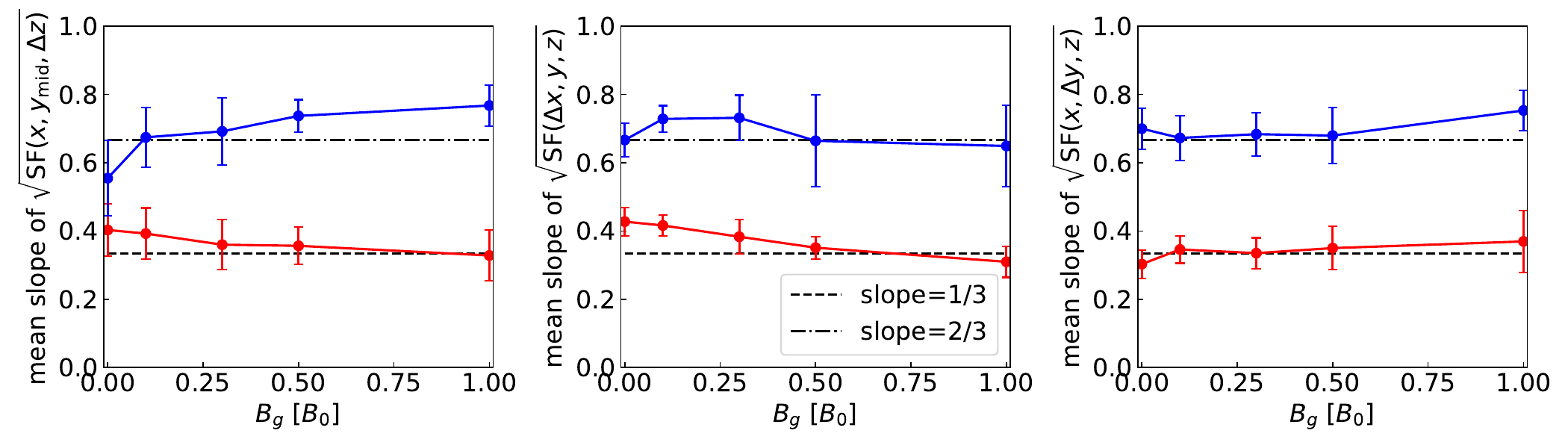}
        \caption{Mean slopes of $\sqrt{{\rm SF}({x,y_{\rm mid}, \Delta z)}}$ (left), $\sqrt{{\rm SF}(\Delta x,y,z)}$ (middle), $\sqrt{{\rm SF}( x,\Delta y,z)}$ (right) as a function of the guide field strength. The slope is fitted over separations in the range 2–30$\, d_e$. The black dashed line indicates the slope of 1/3, while the black dotted-dashed line indicates the slope $2/3$. The error bars show the one sigma standard deviations of the fitted parameters (estimated from the covariance matrix of the least-squares fit). The structure functions are computed at time $t=5.625 \, L/c$. The reconnection region is defined as the volume where both particle populations contribute at least 10\% to the local plasma density.} 
    \label{fig:sf_slope_005_th10}
\end{figure*}

\begin{figure*}
\centering
\includegraphics[width=0.99\linewidth]{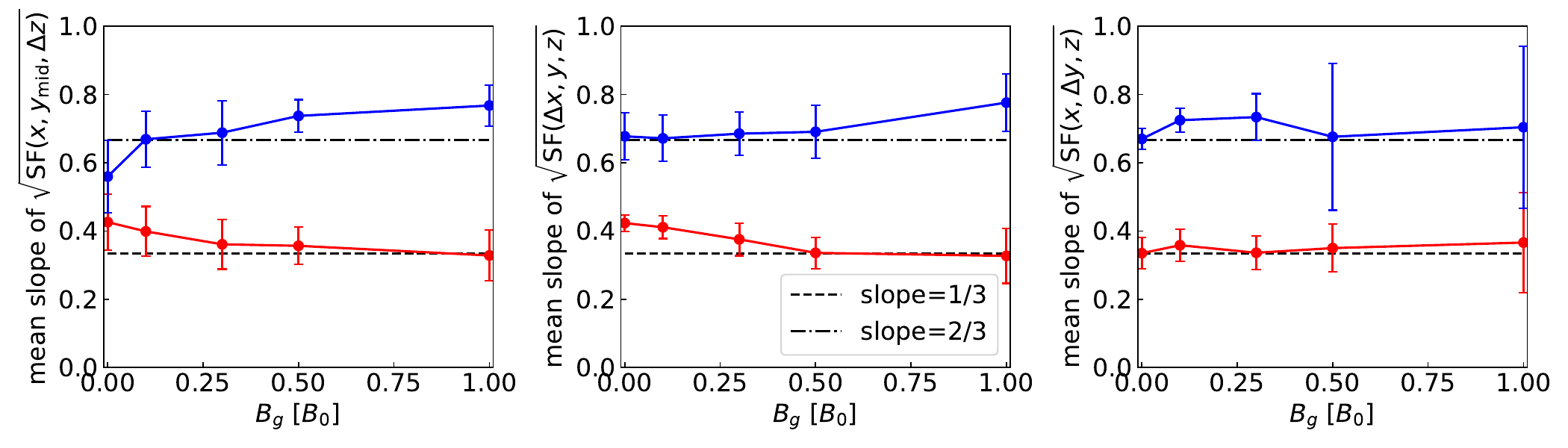}
        \caption{Same as Fig.~\ref{fig:sf_slope_005_th10}, but for the reconnection region defined as the volume where both particle populations contribute at least 20\% to the local plasma density.} 
    \label{fig:sf_slope_005_th20}
\end{figure*}

In Figs.~\ref{fig:sf_slope_004} and \ref{fig:sf_slope_006}, we show the mean slopes of $\sqrt{{\rm SF}(x, y_{\rm mid}, \Delta z)}$, $\sqrt{{\rm SF}(\Delta x, y, z)}$, and $\sqrt{{\rm SF}(x, \Delta y, z)}$ as a function of the guide-field strength. The trend generally agrees with that observed in Fig.~\ref{fig:sf_slope}. In Figs.~\ref{fig:sf_slope_005_th10} and \ref{fig:sf_slope_005_th20}, we show the mean slopes, for reconnection regions defined as the volume where both particle populations contribute at least 10\% and 20\% to the local plasma density, respectively, at $t=5.625\,L/c$. 

\begin{figure*}
\centering
\includegraphics[width=0.93\linewidth]{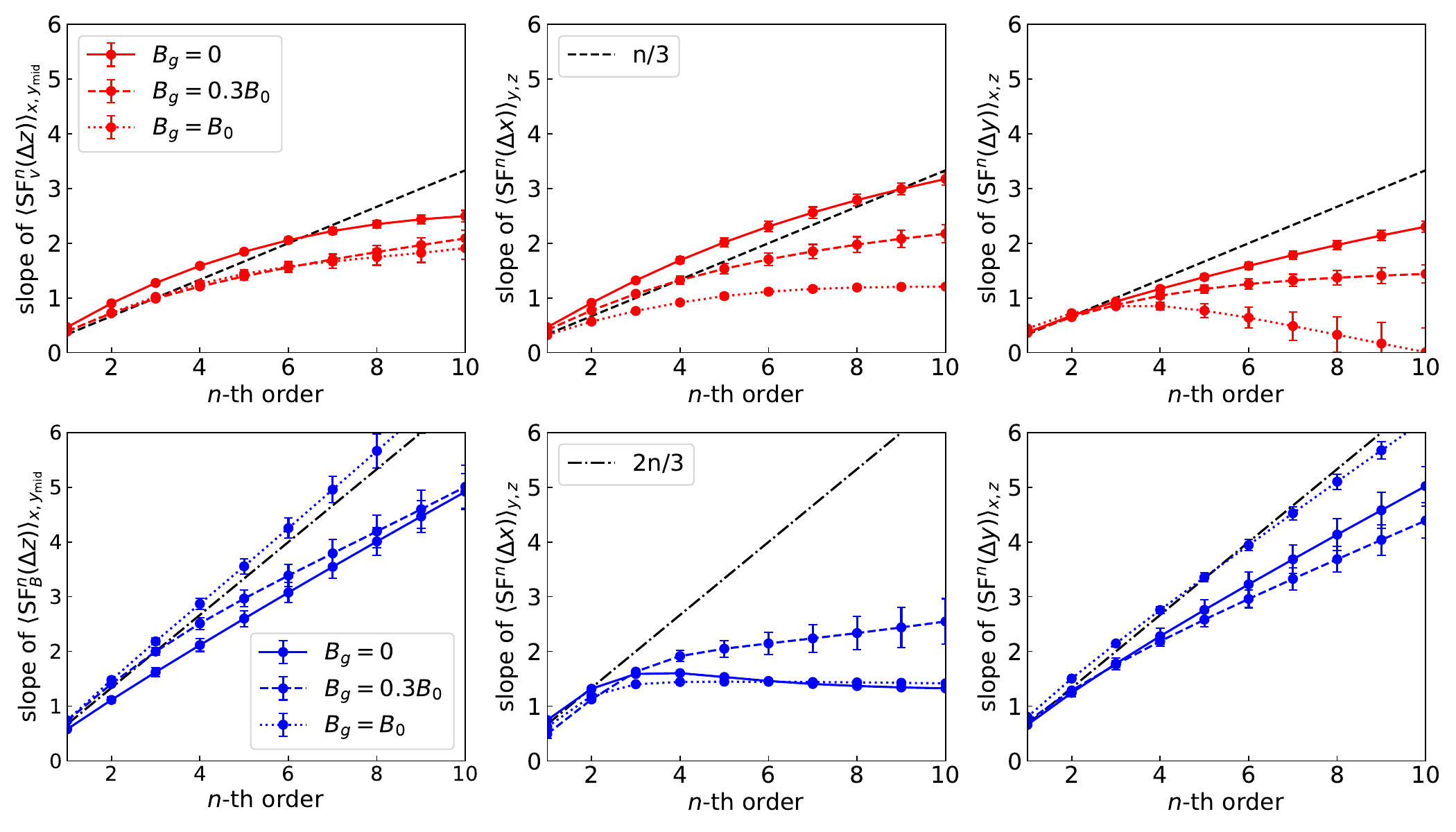}
        \caption{Slopes of the $n$-th order structure functions: $\langle{\rm SF}^n(\Delta z)\rangle_{x,y_{\rm mid}}$ (left), $\langle{\rm SF}^n(\Delta y)\rangle_{x,z}$ (middle), $\langle{\rm SF}^n(\Delta x)\rangle_{y,z}$ (right). The slope is fitted over separations in the range 2–30$\, d_e$. Intermittency is reflected in departures from linear, self-similar scaling. The black dashed line indicates the expected $n/3$ for self-similar Kolmogorov-type fluctuations without intermittency, while the black dotted-dashed line indicates the slope $2n/3$. The error bars show the one sigma standard deviations of the fitted parameters. The structure functions are computed at time $t=4.5 \, L/c$.}
    \label{fig:sf_nth_004}
\end{figure*}

\begin{figure*}
\centering
\includegraphics[width=0.93\linewidth]{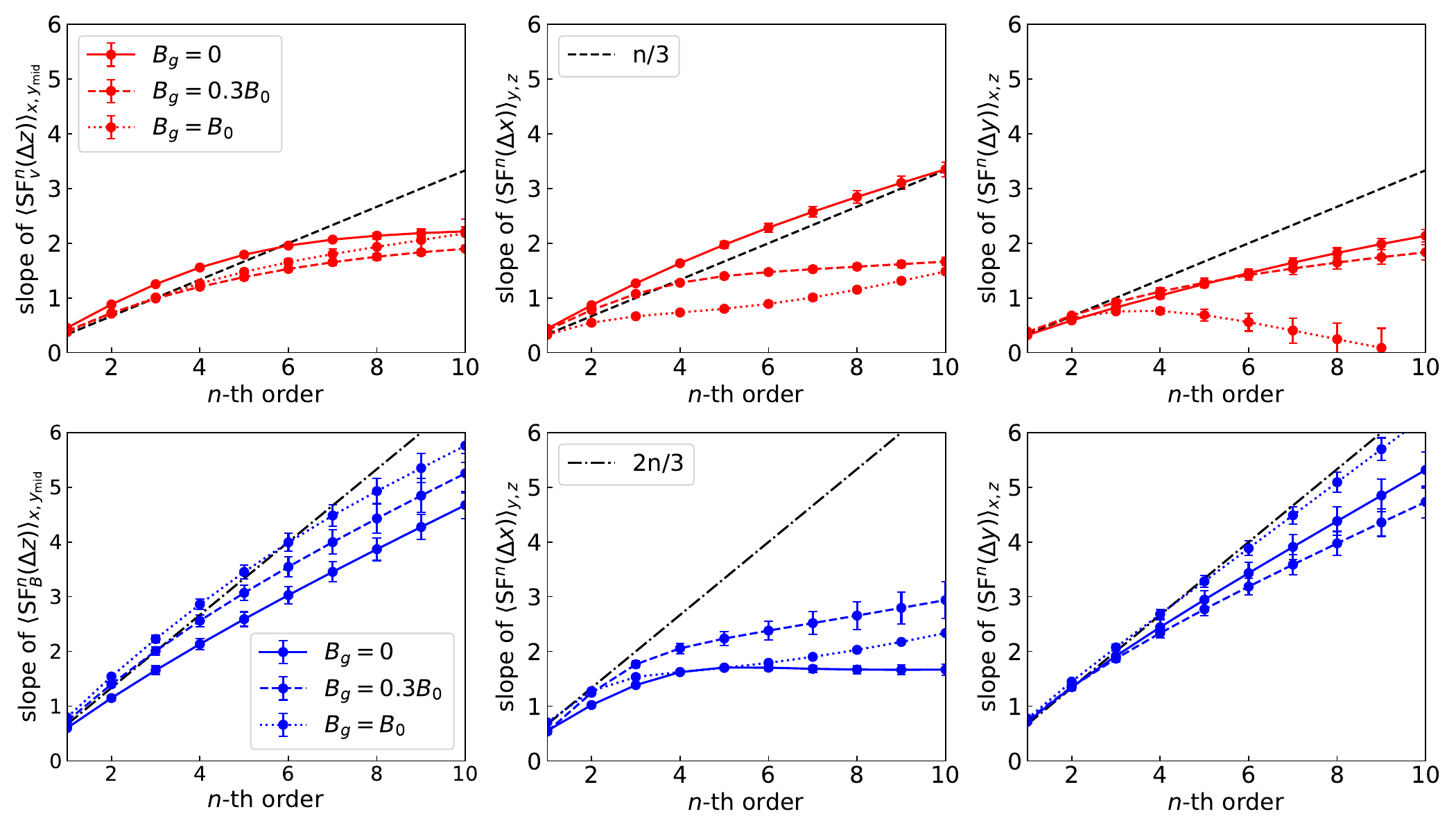}
        \caption{Same as Fig.~\ref{fig:sf_nth_004}, but for the time $t=6.75 \, L/c$.}
    \label{fig:sf_nth_006}
\end{figure*}

In Figs.~\ref{fig:sf_nth_004} and \ref{fig:sf_nth_006}, we present the $n$-th order structure function slopes for orders $n=1$ to $10$ along the out-of-plane ($z$), inflow ($y$), and outflow ($x$) directions, at $t=4.5\,L/c$ and $t=6.75\,L/c$, respectively. Consistent with Fig.~\ref{fig:sf_nth}, we find that the strongest intermittency occurs in the magnetic field along the outflow ($x$) direction, where it exceeds that of the velocity field. The inflow ($y$) direction exhibits weaker intermittency than the velocity field, while the guide-field ($z$) direction is comparable. 

\section{Effect of global magnetic field and velocity structures}
To isolate the turbulent fluctuations from the large-scale reconnection flow and field, we remove two background components before evaluating the structure functions. Based on the statistical homogeneity along the guide-field direction $z$, which has periodic boundary conditions, we subtract the $z$-averaged profile from each velocity and magnetic-field component. This procedure simultaneously removes the reconnection exhaust profile, the inflow profile, and any residual large-scale magnetic structure associated with the mean reconnection geometry (e.g., the reversing in-plane field and the mean out-of-plane component), while preserving the $z$-dependent fluctuations on all scales. 

\begin{figure*}
\centering
\includegraphics[width=0.99\linewidth]{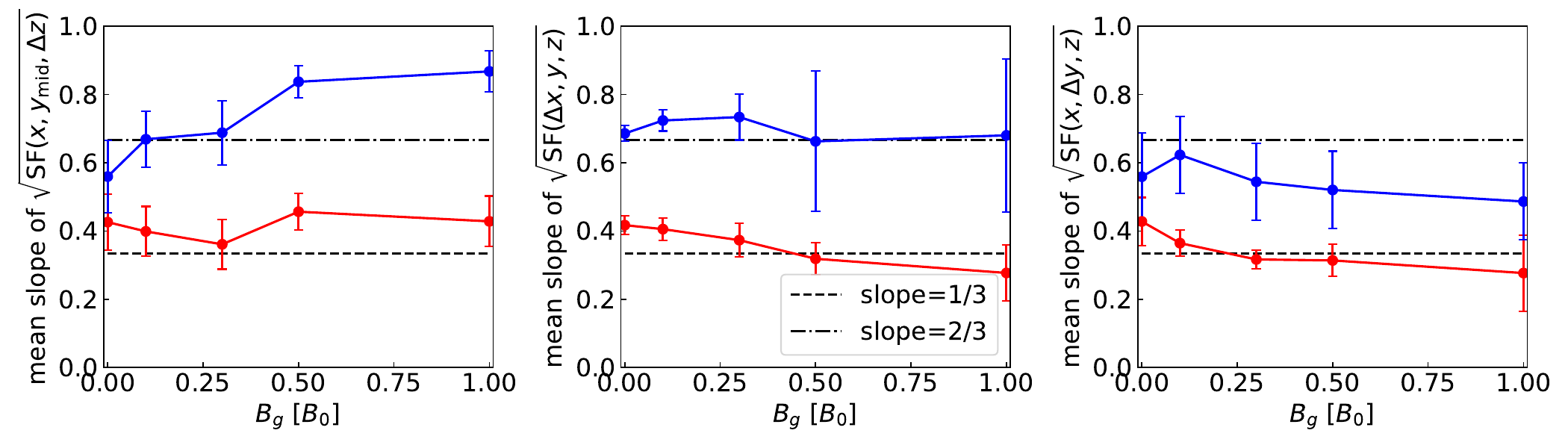}
        \caption{Mean slopes of $\sqrt{{\rm SF}({x,y_{\rm mid}, \Delta z)}}$ (left), $\sqrt{{\rm SF}(\Delta x,y,z)}$ (middle), $\sqrt{{\rm SF}( x,\Delta y,z)}$ (right) as a function of the guide field strength using the mean-subtracted magnetic field and velocity. The slope is fitted over separations in the range 2–30$\, d_e$. The black dashed line indicates the slope of 1/3, while the black dotted-dashed line indicates the slope $2/3$. The error bars show the one sigma standard deviations of the fitted parameters (estimated from the covariance matrix of the least-squares fit). The structure functions are computed at time $t=5.625 \, L/c$.  } 
    \label{fig:sf_slope_005_back}
\end{figure*}

In Fig.~\ref{fig:sf_slope_005_back}, we show the mean slopes of $\sqrt{{\rm SF}(x, y_{\rm mid}, \Delta z)}$, $\sqrt{{\rm SF}(\Delta x, y, z)}$, and $\sqrt{{\rm SF}(x, \Delta y, z)}$ as a function of the guide-field strength using the mean-subtracted magnetic field and velocity.

\section{PDFs of the velocity and magnetic field fluctuations increment}

To characterize the scale-dependent non-Gaussianity of the velocity and magnetic field fluctuations, we compute the probability distribution functions (PDFs) of the increments normalized by their own standard deviation along three directions: $\Delta z$ (guide-field direction, sampled at the current-sheet midplane using periodic boundary conditions), $\Delta x$ (outflow direction), and $\Delta y$ (inflow direction):
\begin{equation}
\begin{aligned}
\label{eq.sf}
\delta \pmb{f}(x,y,\Delta z)&=\pmb{f}(x,y_{\rm mid},z)-\pmb{f}(x,y_{\rm mid},z+\Delta z),\\
\delta \pmb{f}(x,\Delta y,z)&=\pmb{f}(x,y,z)-\pmb{f}(x,y+\Delta y,z),\\
\delta \pmb{f}(\Delta x,y,z)&=\pmb{f}(x,y,z)-\pmb{f}(x+\Delta x,y,z),
\end{aligned}
\end{equation}
, where $\pmb{f} = \pmb{v}$ or $\pmb{f} = \pmb{B}$ denotes the velocity or magnetic field, respectively. We use $r$ to represent the spatial lag $\Delta z$, $\Delta y$, or $\Delta x$. 

\begin{figure*}
\centering
\includegraphics[width=0.99\linewidth]{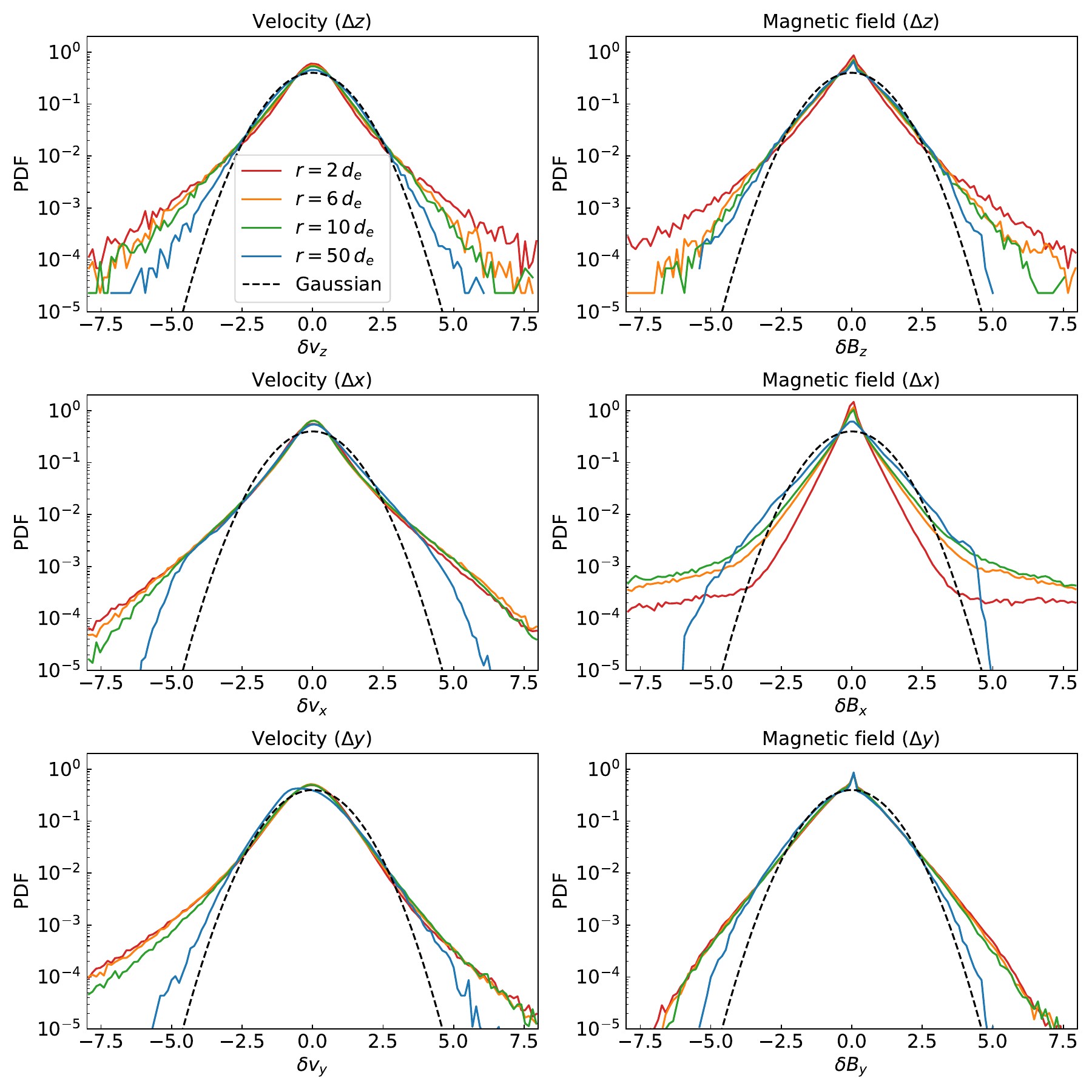}
        \caption{Normalized PDFs of velocity (left) or magnetic field (right) increments at spatial lags $r = 2$, $6$, $10$, and $50\,d_e$ for the $B_g = 0$ simulation at $t=5.625L/c$. Rows from top to bottom correspond to separations along the guide-field direction ($\Delta z$), the outflow direction ($\Delta x$), and the inflow direction ($\Delta y$). The black dashed curve in each panel is the unit Gaussian $\mathcal{N}(0,1)$. }
    \label{fig:pdf_l_005}
\end{figure*}

Fig.~\ref{fig:pdf_l_005} presents the standardized PDFs on semi-logarithmic axes for lags $r = 2$, $6$, $10$, and $50\,d_e$ in the $B_g = 0$ simulation, with the unit Gaussian $\mathcal{N}(0,1)$ shown as a dashed reference curve. At the largest lag ($r = 50\,d_e$), the distributions are close to Gaussian, whereas at progressively smaller lags, the PDF tails become increasingly pronounced — extending well beyond the Gaussian envelope — indicating the presence of intense, localized fluctuations characteristic of intermittency. This trend is evident in all three directions and for both the velocity and the magnetic field, confirming that the turbulence driven by reconnection is intrinsically intermittent.

\begin{figure*}
\centering
\includegraphics[width=0.99\linewidth]{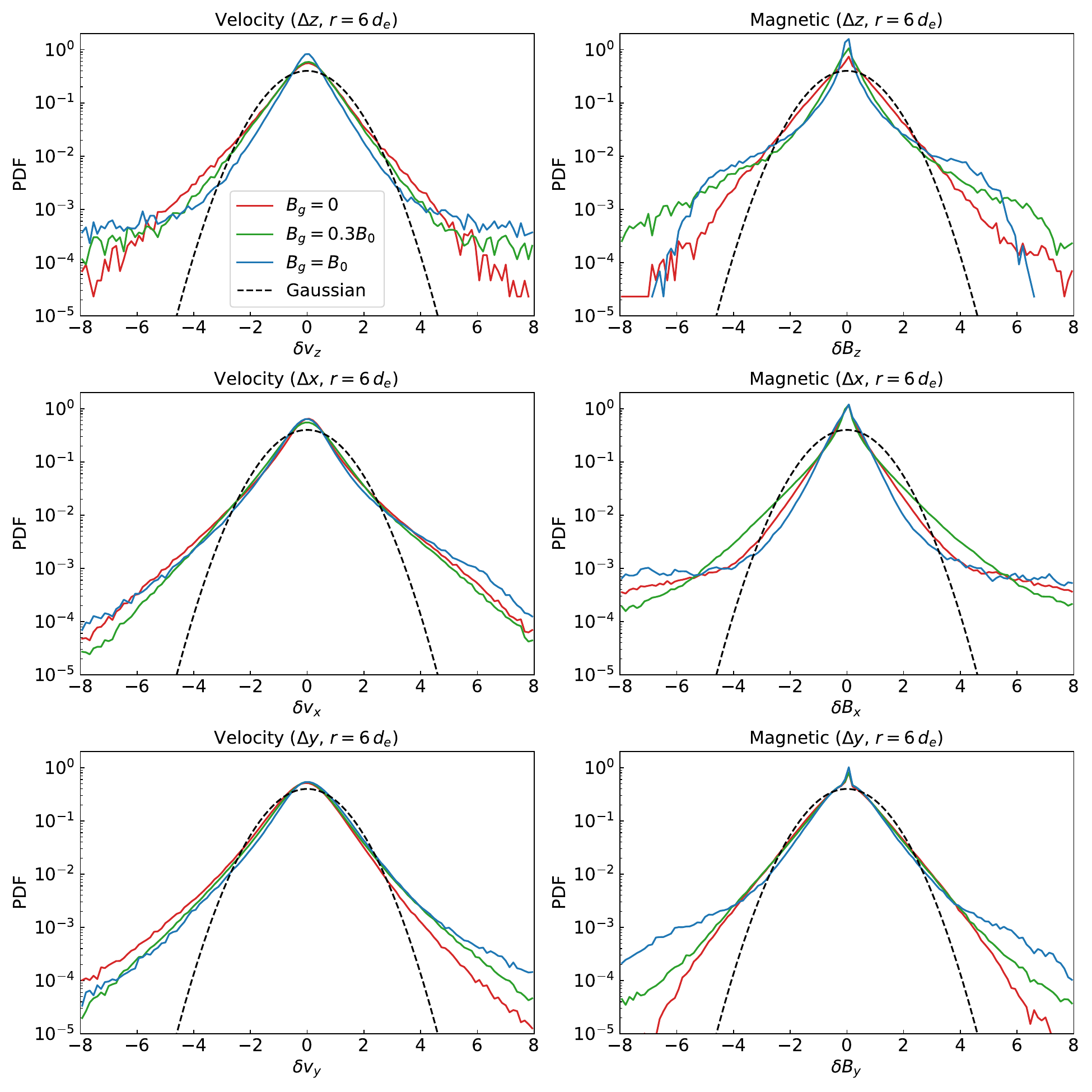}
        \caption{Comparison of normalized increment PDFs across guide-field strengths $B_g = 0$ (red), $0.3\,B_0$ (green), and $B_0$ (blue) at a fixed small lag $r = 6\,d_e$. The panel layout is the same as in Fig.~\ref{fig:pdf_l_005}: the left column shows four-velocity increments and the right column shows magnetic field increments, with rows corresponding to $\Delta z$, $\Delta x$, and $\Delta y$ separations. The black dashed curve is the unit Gaussian.}
    \label{fig:pdf_b_005}
\end{figure*}

To assess how the guide-field strength $B_g$ affects the degree of intermittency, we fix a small spatial lag $r = 6\,d_e$ and compare the standardized increment PDFs across three simulations with $B_g = 0$, $0.3\,B_0$, and $B_0$, where $B_0$ is the reconnecting field strength. Fig.~\ref{fig:pdf_b_005} shows the resulting PDFs for the velocity and magnetic field increments along the same three directions as in Fig.~\ref{fig:pdf_l_005}. At this fixed small scale, differences in the tail directly reflect differences in intermittency: a broader, heavier-tailed PDF indicates a higher probability of extreme increments and hence stronger intermittency. 

\begin{figure*}
\centering
\includegraphics[width=0.99\linewidth]{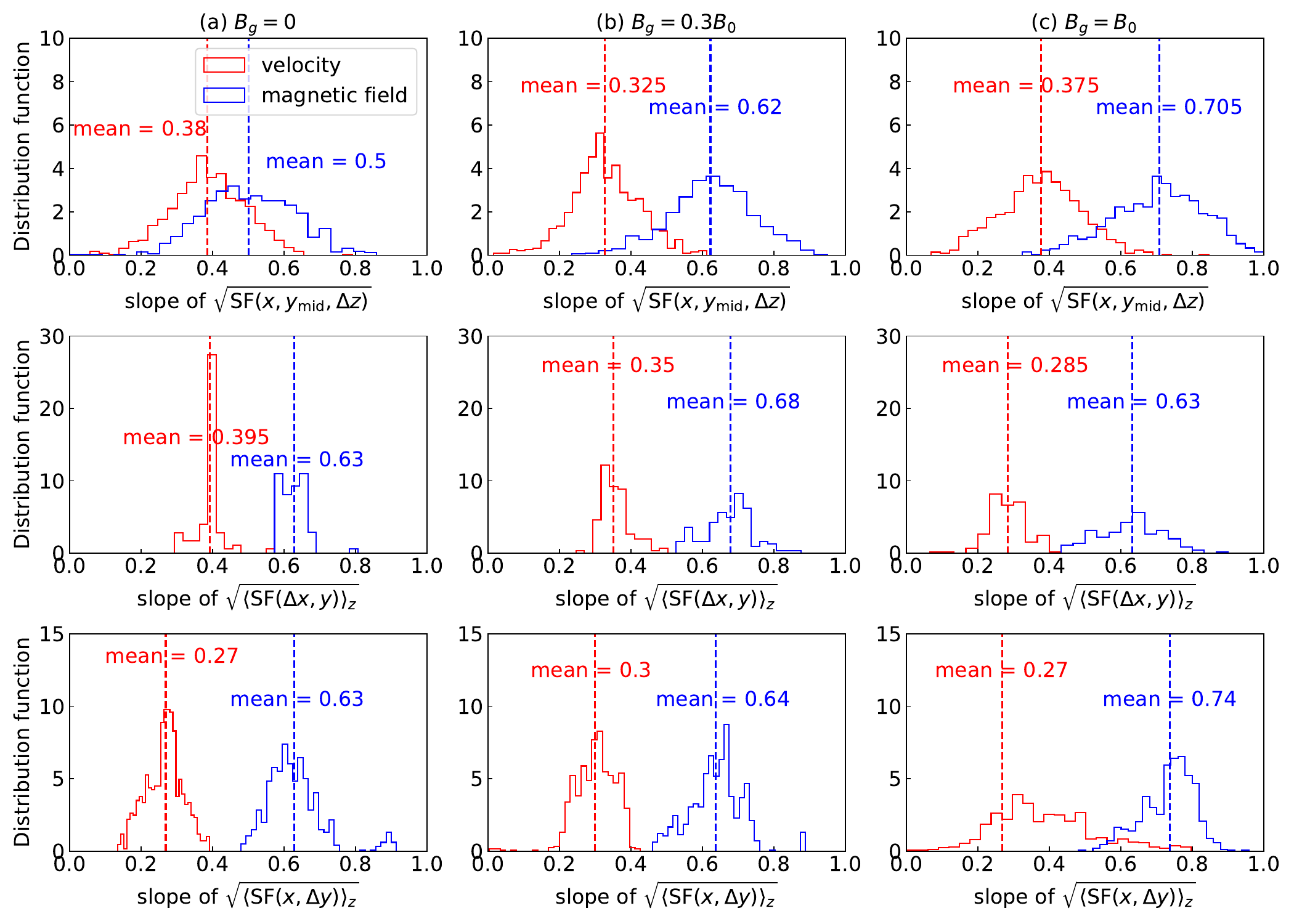}
        \caption{The distribution of the ${\rm SF}(x, y_{\rm mind},\Delta z)$ (top), $\langle{\rm SF}(\Delta x, y)\rangle_z$ (middle), and $\langle{\rm SF}(x, \Delta y)\rangle_z$'s (bottom) slope. The slope is fitted over separations in the range 6 – 30$d_e$ using the simulation at $t=5.625L/c$. The red and blue dashed lines represent the mean slopes of the velocity and magnetic field structure functions, respectively.}
    \label{fig:pdf_b_005}
\end{figure*}



\newpage
\bibliography{sample631,blob,araa}{}
\bibliographystyle{aasjournal}



\end{document}